\documentclass[twocolumn]{aastex63}
\usepackage[utf8]{inputenc}
\usepackage{amsmath}
\usepackage{booktabs}
\usepackage{multirow}
\usepackage{epsfig,lineno}
\usepackage{amsmath,hyperref}
\usepackage{graphicx}
\usepackage{natbib}
\usepackage{soul}

\renewcommand{\vec}{\boldsymbol}
\shortauthors{Lewis et. al.}

\begin{document}

\newcommand\rev[1]{#1} 

\newcommand{\mathcolorbox}[2]{\colorbox{#1}{$\displaystyle #2$}}
\newcommand{\revm}[1]{#1} 

\title{Speckle Space-Time Covariance in High-Contrast Imaging}
\correspondingauthor{Briley Lewis}
\email{blewis@astro.ucla.edu}

\author[0000-0002-8984-4319]{Briley Lewis}
\affiliation{Department of Physics and Astronomy, UCLA, Los Angeles, CA 90024 USA}

\author[0000-0002-0176-8973]{Michael P. Fitzgerald}
\affiliation{Department of Physics and Astronomy, UCLA, Los Angeles, CA 90024 USA}

\author[0000-0002-3973-7646]{Rupert H. Dodkins}
\affiliation{Department of Physics, UCSB, Santa Barbara, CA 93106 USA}

\author[0000-0001-5587-845X]{Kristina K. Davis}
\affiliation{Department of Physics, UCSB, Santa Barbara, CA 93106 USA}

\author[0000-0001-8542-3317]{Jonathan Lin}
\affiliation{Department of Physics and Astronomy, UCLA, Los Angeles, CA 90024 USA}

\begin{abstract}

We introduce a new framework for point-spread function (PSF) subtraction based on the spatio-temporal variation of speckle noise in high-contrast imaging data where the sampling timescale is faster than the speckle evolution timescale. One way that space-time covariance arises in the pupil is as atmospheric layers translate across the telescope aperture and create small, time-varying perturbations in the phase of the incoming wavefront. The propagation of this field to the focal plane preserves some of that space-time covariance. To utilize this covariance, our new approach uses a Karhunen-Lo\'eve transform on an image sequence, as opposed to a set of single reference images as in previous applications of Karhunen-Lo\'eve Image Processing (KLIP) for high-contrast imaging. With the recent development of photon-counting detectors, such as microwave kinetic inductance detectors (MKIDs), this technique now has the potential to improve contrast when used as a post-processing step. Preliminary testing on simulated data shows this technique can improve contrast by at least 10--20\% from the original image, with significant potential for further improvement. For certain choices of parameters, this algorithm may provide larger contrast gains than spatial-only KLIP.

\end{abstract}



\section{Introduction} \label{sec:intro}

Direct imaging of exoplanets is a challenging endeavor, given the extreme contrasts that must be achieved to detect faint planets. Although significant starlight suppression can be achieved through optics and instrumentation, such as coronagraphs, adaptive optics (AO) systems, interferometers, and more, that alone is insufficient to detect analogs of planets in our solar system \citep{oppenheimer2009high,guyon2005limits}. Improving contrast expands the space of the types of planets that can be directly detected and characterized. 

Existing instruments, such as the Gemini Planet Imager \citep{macintosh2008gemini} and VLT's SPHERE \citep{beuzit2019sphere} are able to image giant planets and brown dwarfs, reaching contrasts \rev{(in the astronomical sense, meaning the detectable planet-star flux ratio)} of around $10^{-6}$. This is enabled by a combination of wavefront sensing, control, and post-processing, which reduce the impact of noise by distinguishing between the planet signal and residual noise; this noise arises from uncorrected wavefront aberrations, resulting in quasi-static fluctuations in the focal plane known as ``speckles.'' Generally, these algorithms use the data themselves to create a model of the speckle noise which can then be subtracted from the data to recover the target planet signal in a process known as point-spread function (PSF) subtraction. Previously developed algorithms include LOCI (Locally Optimized Combination of Images; \citet{lafreniere2007new}), KLIP (Karhunen Lo\'eve Image Processing; \citealt{soummer2012KLIP}), and more \citep{gebhard2022half}.  Many directly imaged planet discoveries to date have relied on such algorithms, such as the famous HR 8799 planets \citep{marois2008direct}. 

Improvements to data processing pipelines and methods are one way in which we can push forward and improve contrast for future high-contrast imaging instruments. Other approaches to improving high-contrast imaging methods focus on wavefront sensing and control, such as predictive control techniques, which aim to improve adaptive optics corrections \citep{guyon2017adaptive, guyon2018compute, males2018ground}, and sensor fusion, both currently in development at multiple facilities, including Subaru's SCExAO facility \citep{guyon2017subaru} and Keck Observatory \cite{van2021status,10.1117/12.2560017,jensen2019demonstrating,calvinprep}. Other recent work such as \citet{guyon2017adaptive} focuses on using on Empirical Orthogonal Functions (EOFs), a similar mathematical framework, to analyze spatio-temporal correlations; their work is in the context of predictive control, whereas our work applies to image processing. New advances in detector technology also affect both wavefront sensing and post-processing. High-speed, low-noise detectors will provide multiple opportunities for improvements, including focal-plane wavefront sensing, which eliminates non-common-path wavefront errors \citep{vievard2020focal}. Of particular interest are arrayed photon-counting devices, such as Microwave Kinetic Inductance Detectors (MKIDs) \citep{schlaerth2008millimeter, mazin2012superconducting, meeker2018darkness, walter2020mkid} and Infrared Avalanche Photodiodes (IR APDs) \citep{2018arXiv180704903G,2021SPIE11763E..1JW}. Electron Multiplying CCDs (EMCCDs) are a functional equivalent in the optical \citep{lake2020developments}. Photon arrival times have already been used to distinguish speckles from incoherent signals, such as planets \citep{walter2019stochastic,steiger2021scexao}, and MKIDs have been used for high contrast imaging with the DARKNESS instrument at Palomar \citep{meeker2018darkness} and with MEC, the MKID Exoplanet Camera for high contrast astronomy at Subaru \citep{walter2018mec}. 

This new regime of photon-counting detectors and more advanced adaptive optics presents many opportunities. With the improved temporal resolution of next-generation detectors, we will be able to resolve the spatial and temporal evolution of atmospheric speckles. Some prior work has investigated use of spatio-temporal correlations on longer timescales, such as \citet{mullen2019speckle} and \citet{gebhard2022half}, but this work focuses the shorter timescale changes of atmospheric speckles. 

There is a rich history of theory and measurements of space-time atmospheric speckle behavior in the past decades, which this work builds off of. Since the 1970s--1980s, speckle patterns and intensity distributions have been measured \citep{1981JOSA...71..490D,1978ApOpt..17.3779S,2018arXiv180704903G,1982OptCo..41...79O}, demonstrating agreement with models based in Rician statistics \citep{2001OptEn..40.2690C,1999ApOpt..38..766C} and the importance of speckles as the limiting noise source in the high-contrast regime \citep{1999PASP..111..587R}. The space-time covariance was even directly measured in \citet{1981JOSA...71..490D}, indicating that speckle boiling has a directionality related to turbulence. Speckle intensity patterns have been modeled as a modified Rician distribution \citep{2004ApJ...612L..85A,2010JOSAA..27A..64G}, and speckle lifetimes have been constrained through models and direct measurements \citep{1986JOSAA...3.1001A, 1991A&A...243..553V, 1993JMOp...40.2381G}. In fact, models of speckle boiling directly relate the lifetime of speckles to atmospheric parameters related to wind and turbulence, as in \citet{1982JOpt...13..263R}, estimating speckle lifetimes on the order of tens of milliseconds. 

\rev{This work is a new addition to the variety of time-domain algorithms that have been developed in recent years. For example, the PACO algorithm uses temporal information from the background fluctuations of Angular Differential Imaging data} \citep{flasseur2018exoplanet}\rev{, and the TRAP algorithm uses temporal information of the speckle pattern to improve contrast specifically at close separations} \citep{samland2021trap}. \rev{Another algorithm, from} \citet{gebhard2022half}, \rev{uses half-sibling regression on time-series data. These are all examples of the possibilities for temporal information in post-processing, in addition to the AO control improvements described earlier.}

In this work, we aim for a second-order characterization of the statistical behavior of atmospheric speckles in the high-contrast regime, described by the space-time covariance, which we then leverage for improving contrast in post-processing with the eventual goal of improving exoplanet detection capabilities. As previously mentioned, this goal is not without its challenges --- with kHz readouts, these detectors can produce large datasets and lead to computationally intensive post-processing methods. While developing this new technique, we must also contend with data storage and computational limitations. 

In this paper, we first provide an analytical justification for the existence of these covariances in the high-contrast regime, observe their occurrence in test simulations \rev{focusing on millisecond time sampling}, and then present an initial algorithm to exploit these covariances for PSF subtraction. Specifically, we are testing this algorithm in a regime dominated by atmospheric speckles at short exposures (where the timescale of our exposures is short compared to that of changes in atmospheric residual wavefront error, so atmosphere is essentially frozen). 

Here in Section \ref{sec:theory}, we describe the process of baseline Karhunen-Lo\'eve Image Processing (KLIP), the origins of space-time speckle covariances, and the extension of KLIP to space-time covariances. Following, in Section \ref{sec:methods}, we describe the models used to create datasets for initial testing of this processing framework. Section \ref{sec:results} presents results of this new algorithm implemented on simulated data. Finally, in Sections \ref{sec:discussion} and \ref{sec:conclusions}, we discuss the promise of this new technique, as well as its current challenges/limitations and future work.

\section{Space-Time Covariance Theory} \label{sec:theory}

Speckles can limit contrast, but can also be subtracted to some extent to improve contrast. One of the most successful post-processing algorithms has been KLIP, described in Section \ref{subsec:KLIP}, which exploits spatial correlations in long-exposure images. We motivate our extension of this technique to include space-time correlations on shorter timescales in Section \ref{subsec:sttheory} by describing how these correlations arise in imaging through the atmosphere. This extension of KLIP, referred to as space-time KLIP or stKLIP, is demonstrated in Section \ref{subsec:stKLIP}, exploiting spatio-temporal correlations between short-exposure images.

\subsection{Karhunen-Lo\'eve Image Processing} \label{subsec:KLIP}

Karhunen-Lo\'eve Image Processing is a data processing technique \rev{that uses} principle component analysis (PCA)\rev{,} where data are represented by a linear combination of orthogonal functions. In high-contrast imaging, KLIP is used to build a model, used for PSF subtraction, that accounts for spatial correlations between speckles and other PSF features, first described in \citet{soummer2012KLIP}. This technique takes advantage of spatial covariances of the speckles in the image, because strong correlations exist in high eigenvalue modes and can be suppressed. This is a data-driven approach, which uses available information from the data itself to provide an approximation of the noise, by using a subset of the data as ``reference images'' from which to build the model of the noise while using another subset of the data as the ``target image'' for PSF subtraction. 

To increase readability, all variables for the following mathematics are described in Appendix A. As described in \citet{soummer2012KLIP}, we assume we observe a point spread function $T(k)$, where $k$ is the pixel index, that contains the stellar point spread function $I_\psi(k)$ and may also contain some faint astronomical signal of interest $A(k)$. Therefore, our target image can be described as
\begin{equation}
    T(k) = I_\psi(k) + \epsilon A(k),
\end{equation}
where $\epsilon$ is 0 when there is no astronomical signal of interest, or 1 if there is. The goal of PSF subtraction is therefore to recreate $I_\psi(k)$ in order to isolate $A(k)$. Without an infinite number of references, though, we cannot exactly infer $I_\psi(k)$; instead, we approximate the PSF $\hat{I_\psi}(k)$. \rev{For consistency in our notation, herein we represent} $\revm{T(k)}$, $\revm{A(k)}$,\rev{ and }$\revm{\hat{I_\psi}(k)}$\rev{ as vectors }$\revm{\vec{t}}$, $\revm{\vec{a}}$\rev{ and }$\revm{\vec{\hat{\psi}}}$\rev{ respectively.}

In order to approximate $\revm{\vec{\hat{\psi}}}$, KLIP computes a Karhunen-Lo\'eve Transform based on the covariance matrix of the mean-subtracted\rev{ }reference images. 

\rev{A sequence of reference images are first unraveled into one-dimensional vectors, each as} $\revm{\vec{r}}$. \rev{Note: henceforth vectors are denoted as bold, matrices with uppercase variables and subscript matrix elements. These image vectors }$\revm{\vec{r}}$\rev{ are then stacked into an }$\revm{n_p \times n_{i}}$\rev{ matrix }$\revm{R}$\rev{, where }$\revm{n_{p} = n_x \times n_y}$\rev{ and }$\revm{n_i}$\rev{ is the number of images, as follows:} 

\begin{equation}\label{refimg}
    \revm{R = }\begin{bmatrix}
R_{1,1} & R_{1,2} & \hdots & R_{1,n_i}\\
R_{2,1} & R_{2,2} & \hdots & R_{2,n_i}\\
\vdots & \vdots & \ddots & \vdots\\
R_{n_p,1} & R_{n_p,2} & \hdots & R_{n_p,n_i}
\end{bmatrix}
\end{equation}

We then subtract the mean image of the set \rev{(summing over the matrix columns)} from the reference set $R$, in order to produce a set of mean-subtracted images $M$ to use throughout the process of KLIP:

\begin{equation}\label{eq:ms}
   \vec{x}_i = \frac{1}{n_i}\sum_{j=1}^{n_i} R_{i,j}
\end{equation}
\begin{equation}
   M_{i,j} = R_{i,j} - \vec{x}_i
\end{equation}

The resulting covariance matrix (\ref{eq:klcov}) $C$ has size $n_{p} \times n_{p}$. 

\begin{equation}\label{eq:klcov}
    C = MM^T = \begin{bmatrix}
C_{1,1} & C_{1,2} & \hdots & C_{1,n_p}\\
C_{2,1} & C_{2,2} & \hdots & C_{2,n_p}\\
\vdots & \vdots & \ddots & \vdots\\
C_{n_p,0} & C_{n_p,1} & \hdots & C_{n_p,n_p}
\end{bmatrix}
\end{equation}

Note: in practice, this implementation is computationally expensive, so the covariance is instead \rev{often} computed in image space on $n_{i}$ by $n_{i}$ images and then re-projected into pixel space, as is done in the \citet{soummer2012KLIP} implementation. \rev{The ideal implementation depends on which dimension is larger / more computationally expensive, e.g.} \citet{2021AJ....161..166L}. In this work, the mathematics for KLIP and stKLIP, as written here, will be in pixel space.

An eigendecomposition of the covariance matrix $C$\rev{, mathematically described as solutions to the equation}
\begin{equation}\label{eigh}
    C\vec{v}_j = \lambda_j \vec{v}_j,
\end{equation}
\rev{with}
\begin{equation}
    \lambda_1 > \lambda_2 > \lambda_3 > \hdots \lambda_{n_p},
\end{equation}
produces a length $n_p$ vector of eigenvalues ($\vec{\lambda}$) and size $n_p \times n_p$ (or $n_m \times n_p$ if fewer than $n_p$ eigenvectors/modes are used) matrix of eigenvectors/eigenimages ($V$) containing $n_m$ rows of individual eigenvectors $\vec{v}$ each of length $n_p$, such that $V_{i,j} = ({\vec{v}_j})_i$. 
\begin{equation}
    V = \begin{bmatrix}
V_{1,1} & V_{1,2} & \hdots & V_{1,n_p}\\
V_{2,1} & V_{2,2} & \hdots & V_{2,n_p}\\
\vdots & \vdots & \ddots & \vdots\\
V_{n_m,1} & V_{n_m,2} & \hdots & V_{n_m,n_p}
\end{bmatrix}
\end{equation}
The eigenvalues order the eigenimages by their importance to rebuilding the original image and are used to construct the basis of the new subspace of greatest variation onto which we project our images. Assuming the vectors are sorted by decreasing eigenvalue, the first coordinate corresponds to the direction of greatest variation. The lowest-order (first coordinate) eigenimages are selected to represent $\vec{\hat{\psi}}$, while leaving the high-order terms to hopefully contain our astrophysical signal. 

We select a given number $n_{\rm m}$ of the eigenimages as our number of modes of variation. The inner product of \rev{the matrix} of eigenvectors $V$ with the one-dimensional vector of the target image $\vec{t}$ \rev{(which has length $n_p$), is described mathematically as}

\begin{equation}\label{target}
    \vec{t} = \begin{bmatrix}
t_{1}\\
t_{2} \\
\vdots \\
t_{n_p}
\end{bmatrix}
\end{equation}

\begin{equation}\label{eq:inner}
    \vec{q} = V \cdot \vec{t} = \begin{bmatrix}
q_{1}\\
q_{2} \\
\vdots \\
q_{n_m}
\end{bmatrix}
\end{equation}
\rev{and} creates a vector of coefficients $\vec{q}$ of length $n_{\rm m}$ --- \rev{each of} these can be thought of as how much of each mode (or each eigenvector, $v_j$) is in the image, or equivalently, the coordinates in the new rotated principle axis space. 

Lastly, we can project back into our original pixel space by taking the product of this vector of coefficients with the chosen eigenvectors, recovering a vector of length $n_{\rm p}$, the same as our target image\rev{:} 
\begin{equation}\label{eq:proj}
    \vec{\hat{\psi}} = \vec{q}^T \cdot V
\end{equation}The resulting array is our image projected into the subspace of greatest variation, \rev{an} estimation of the original PSF $\vec{\hat{\psi}}$, and what we will subtract from our target image for PSF subtraction. Note that the tuneable parameter here is the number of eigenvectors used in the basis (the number of ``modes'').

The planet signal is also projected onto a distribution of these modes, and it is assumed that the planet signal is primarily projected onto modes with lower eigenvalue. However, as we subtract more modes, the projection of the planet onto these modes is also subtracted. Therefore, a larger number of modes might lead to oversubtraction of a planet signal, but too few may not sufficiently subtract out the speckle noise. As a result, we must correct for this throughput effect and optimize the number of modes to attain the largest possible contrast gain. 

\subsection{Space-Time Covariances} \label{subsec:sttheory}

Whereas KLIP harnesses spatial covariances of speckle noise, we propose to expand the scope of such projection methods to take advantage of space-time covariances in speckle noise. For bulk flow in a turbulent atmosphere, phase errors in the pupil, from atmospheric disturbances, translate across the telescope with wind motion, resulting in changes in phase and amplitude in the image plane. 
Atmospheric perturbations evolve across a broad set of spatial frequencies.  Since the perturbations at these different spatial frequencies are correlated, we will illustrate that the speckles at the locations that correspond to those spatial frequencies in the image plane will be correlated as well. Similarly to the above section, all variables for the following mathematics are described in Appendix B.

The covariance of intensity in the image plane for points separated in space and time is characterized through the second moment $\langle I(\vec{x}_1,t)I(\vec{x}_2,t-\tau)\rangle$, where $I$ is the intensity in the image. Angle brackets ($\langle \rangle$) denote averaging over a statistical ensemble. Suppose we have a perfect coronagraph and only phase aberrations are present, ignoring polarization as well as static phase errors, and treating electric field as a scalar.  Also, we presume the phase aberrations are small, a reasonable assumption for the high-contrast imaging limit.  In this case, the pupil amplitude \rev{is}
\begin{equation}
\Psi_{\rm pup}(\vec{u},t) = P(\vec{u})e^{i\phi(\vec{u},t)},
\end{equation}
\rev{approximated as}
\begin{equation}
\Psi_{\rm pup}(\vec{u},t) \approx [1+i\phi(\vec{u},t)]P(\vec{u}),\label{eq:approx}
\end{equation}
where $P(\vec{u})$ is the pupil function, $\phi$ is the phase, and $\vec{u}$ is the coordinate in the pupil plane ($\vec{x}$ is the coordinate in the focal plane, related by a Fourier transform). \rev{It is worth noting that departure from this assumption of linearity may affect results.} The amplitude in the focal plane is
\begin{eqnarray} 
\Psi_{\rm foc}(\vec{x},t) &=& \mathcal{F}\left\{P(\vec{u})\right\} + i\mathcal{F}\left\{\phi(\vec{u},t)P(\vec{u})\right\}, \label{two}\\
&=& C(\vec{x})+S_\phi(\vec{x},t).\label{eq:three}
\end{eqnarray}
$C(\vec{x})$ is the spatially coherent part of the wavefront, and $S_\phi(\vec{x},t)$ comes from phase aberrations -- $S_\phi(\vec{x},t)$ corresponds to the ``speckles'' we want to remove \citep{aime2004usefulness, roddier1982origin}. In the case of a perfect coronagraph, $C(\vec{x})=0$ and the intensity in the image is only due to phase aberrations, and can be expressed as
\begin{eqnarray}
I(\vec{x},t) &=& |\Psi_{\rm foc}(\vec{x},t)|^2, \\
&=& |S_\phi(\vec{x},t)|^2, \\
&=& |\mathcal{F}\left\{\phi(\vec{u},t)P(\vec{u})\right\}|^2.
\end{eqnarray}
The covariance of the intensity is
\begin{equation}
\langle I(\vec{x}_1,t)I(\vec{x}_2,t-\tau)\rangle = \langle|S_\phi(\vec{x}_1,t)S_\phi(\vec{x}_2,t-\tau)|^2\rangle.
\end{equation}
If we assume (complex) Gaussian statistics for $S_\phi$~\citep{soummer2007speckle}, then by Wick's theorem~\citep[e.g.][]{fassino_etal19} we have,
\begin{multline}
\langle I(\vec{x}_1,t)I(\vec{x}_2,t-\tau)\rangle = \\ \langle I(\vec{x}_1,t)\rangle\langle I(\vec{x}_2,t)\rangle + |\langle S_\phi(\vec{x}_1,t)S_\phi^*(\vec{x}_2,t-\tau)\rangle|^2.
\end{multline}
Therefore to compute this covariance, we need the quantity
$\langle S_\phi(\vec{x}_1,t)S_\phi^*(\vec{x}_2,t-\tau)\rangle,$
which is the covariance of the phase-induced aberration in the focal plane.  Accounting for the Fourier relationship between the focal plane aberration $S_\phi$ and the pupil plane phase $\phi$ as in Equations \ref{two} and \ref{eq:three}, we find
\begin{multline}\label{eq:nine}
\langle S_\phi(\vec{x}_1,t)S_\phi^*(\vec{x}_2,t-\tau)\rangle = \\
\int d\vec{u}\int d\vec{\xi} \exp[2\pi i\vec{\xi}\cdot\vec{x}_2-2\pi i\vec{u}\cdot(\vec{x}_1-\vec{x}_2)]\times \\ \langle\phi(\vec{u},t)\phi(\vec{u}+\vec{\xi},t-\tau)\rangle P(\vec{u})P(\vec{u}+\vec{\xi})
\end{multline}
where $\vec{\xi}$ is the coordinate of the displacement in the pupil plane. If $\phi(\vec{u},t)$ is statistically stationary in the pupil plane position $\vec{u}$ (and time), then we can define \rev{the phase covariance function as}
\begin{equation}\label{eq:phase-func}
B_\phi(\vec{\xi},\tau) = \langle\phi(\vec{u},t)\phi(\vec{u}+\vec{\xi},t-\tau)\rangle,
\end{equation}
independent of $\vec{u}$ and $t$. Equation \ref{eq:phase-func} for $B_\phi$ relates space-time covariance in the pupil to space-time covariance in the image, and can be simplified into the Kolmogorov phase covariance function for turbulence with an assumption about time. 

Kolmogorov's theory of turbulence describes a cascade of large scale turbulent motions that dissipate energy onto smaller scales, following a power spectrum described by $\Phi_n(\vec{k}) \propto |\vec{k}|^{-11/3}$, where $\Phi_n$ is the variation in index of refraction and $|\vec{k}|$ is the magnitude of the turbulence \citep{kolmogorov1941dissipation, hickson2008fundamentals}. Fluctuations in density correspond to fluctuations in the index of refraction. These variations in index of refraction lead to differences in path length for the incoming light, creating some of the phase and amplitude error that we observe. However, we assume the timescale of change for this turbulence is generally slow when compared to wind speeds, an assumption known as \rev{Taylor frozen flow} \citep{taylor1938spectrum}. This assumption is valid so long as the turbulent intensity is low compared to the wind speed, generally accepted to be true for astronomical contexts with the possible exception of boundary layer turbulence \citep{bharmal2015frozen}. The turbulence can be thought of then as a ``phase screen'' propagating horizontally across the telescope with the wind. \rev{This phenomenon is described mathematically as} 

\begin{equation} \label{eq:taylor}
    \phi(\vec{u},t) = \phi(\vec{u}-\vec{v}_{\rm wind}\tau,t-\tau)
\end{equation}
which states that the phase structure at one time is related to the phase structure at a different time, just shifted by the wind velocity times the time difference \citep{taylor1938spectrum,hickson2008fundamentals}.

This shows that a single phase screen $\phi(\vec{u},t)$ (which contains Kolmogorov turbulence $\Phi_n$) under Taylor frozen flow is related to a phase screen at a different time $\phi(\vec{u},t-\tau)$ via the wind speed $v_{\rm wind}$. Similarly, we can then say
\begin{equation}\label{eq:kol}
    B_{\phi}(\vec{u},t) = B_{\phi}(\vec{u}-\vec{v}_{\rm wind}\tau,t-\tau).
\end{equation}
This implies the phase \rev{covariance} function at one location and time $B_\phi(\vec{\xi},t)$ in the pupil is related to the phase \rev{covariance} function at that location at a previous time $B_\phi(\vec{\xi},0)$, where $B_\phi(\vec{\xi},0)$ is a covariance related to the Kolmogorov phase covariance function. Since we know the Kolmogorov phase covariance function is non-zero as long as turbulence is present, this demonstrates that the phase \rev{covariance} function at an arbitrary location and time $B_\phi(\vec{\xi},\tau)$ is non-zero. Even if frozen flow is violated, as long as there is non-zero space-time covariance in the pupil, we expect non-zero space-time covariance in the image, as shown in Equation \ref{eq:phase-func}.

Rearranging Equation \ref{eq:nine},
\begin{multline}
\langle S_\phi(\vec{x}_1,t)S_\phi^*(\vec{x}_2,t-\tau)\rangle = \\
\int d\vec{\xi} \exp(2\pi i\vec{\xi}\cdot\vec{x}_2) B_\phi(\vec{\xi},\tau) \\
\int d\vec{u} \exp[-2\pi i\vec{u}\cdot(\vec{x}_1-\vec{x}_2)] P(\vec{u})P(\vec{u}+\vec{\xi}).
\end{multline}
The latter integral is the Fourier transform of the overlap of displaced pupils.  Defining this function,
\begin{equation}
\mathcal{P}(\vec{r},\vec{\xi}) = \int d\vec{u} \exp(-2\pi i\vec{u}\cdot\vec{r}) P(\vec{u})P(\vec{u}+\vec{\xi}),
\end{equation}
we now have the space-time covariance of speckles as the product of the turbulence phase \rev{covariance} function and $\mathcal{P}$, as follows:
\begin{multline}\label{eq:25}
\langle S_\phi(\vec{x}_1,t)S_\phi^*(\vec{x}_2,t-\tau)\rangle = \\ 
\int d\vec{\xi} \exp(2\pi i\vec{\xi}\cdot\vec{x}_2) B_\phi(\vec{\xi},\tau) \mathcal{P}(\vec{x}_1-\vec{x}_2,\vec{\xi}).
\end{multline}

This mathematical framework illustrates how the focal plane covariance is intimately related to pupil plane covariance in the high contrast imaging regime, with a perfect coronagraph and small phase errors. Looking at the overlap of displaced pupils, $\mathcal{P}(\vec{x}_1-\vec{x}_2,\vec{\xi})$, the form of the expression suggests that covariance will be strongest at smaller spatial separations. Similarly, Equation \ref{eq:kol} suggests that covariance will be strongest at smaller temporal separations. Overall, if there is non-zero space time covariance in the pupil plane, then we will have non-zero space time covariance in the focal plane. We will test this further with simulations, as described in Section \ref{sec:methods}.

\subsection{Space-Time KLIP}\label{subsec:stKLIP}

Recall that KLIP improves contrast by projecting away features that are spatially correlated in image sequences. We can extend the framework of KLIP \citep{soummer2012KLIP} to space-time covariances by using an image sequence instead of an image. Note that for the following mathematics we assume discrete time sequences, rather than continuous as in Section \ref{subsec:sttheory} above. \rev{Additionally, we assume regular and continuous time sampling for this implementation; however, this method can be extended easily to block-continuous sampling, which may be useful in future work.} 

All variables for the following mathematics are also described in Appendix A. Baseline KLIP uses an image vector of length $n_{\rm p}$ (number of pixels in image) as its target image and a $n_p \times n_i$ matrix as the set of reference images to determine covariance between pixels, find eigenvectors of covariance, and project out the largest eigenvalue modes from the image. Similarly, space-time KLIP (referred to as stKLIP) uses an image sequence of length $n_s \times n_{\rm p}$ (number of images in the sequence times number of pixels per image), as shown in Equation \ref{stklipmatrix}, to perform those steps. Note that this is transposed compared to KLIP, which uses $n_{\rm p} \times n_s$.

\begin{figure*}
    \centering
    \includegraphics[width=0.9\linewidth]{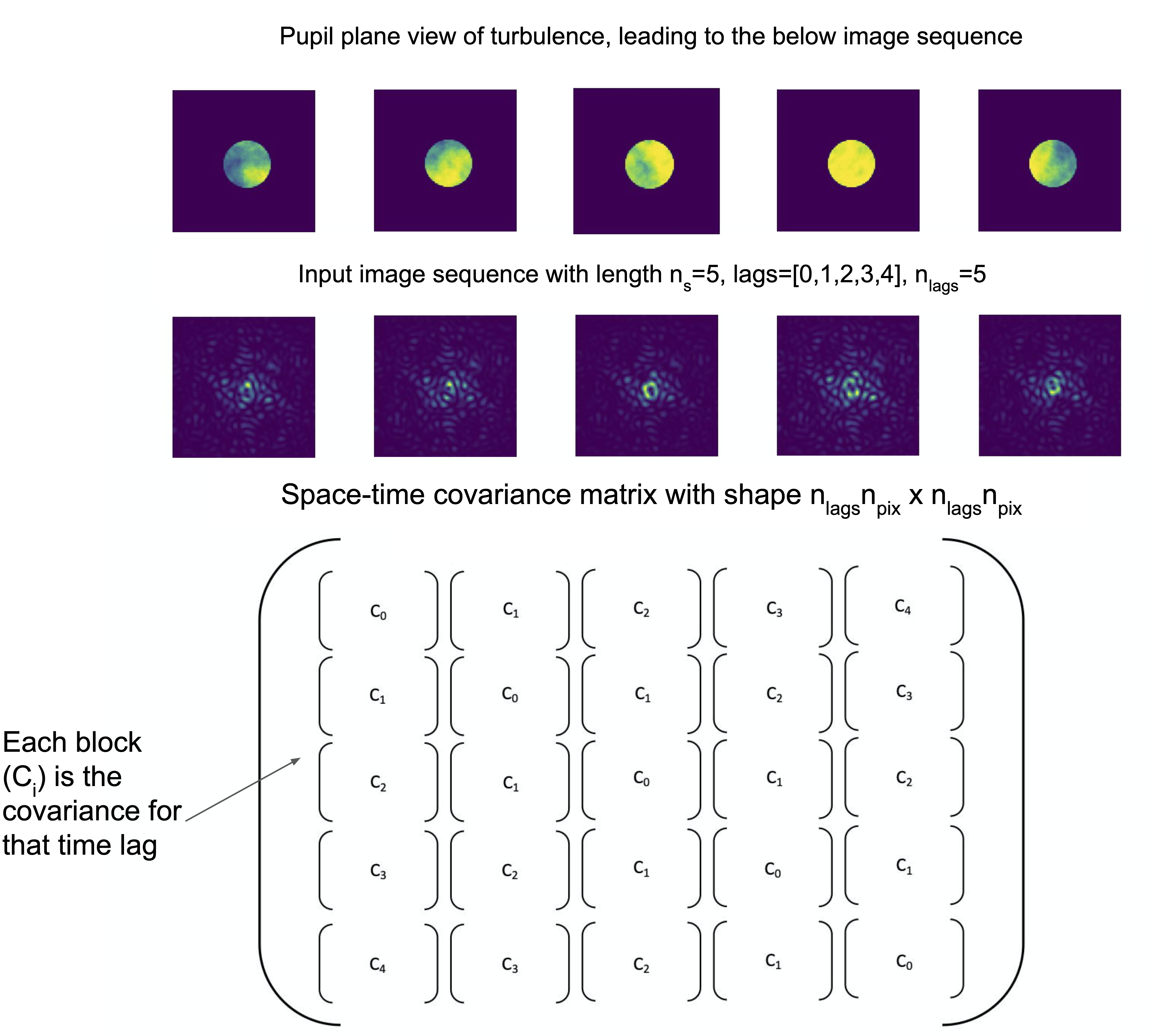}
    \caption{Diagram of stKLIP input sequence setup -- translating phase screens (top) and resulting image sequence (middle) -- with the corresponding block diagonal space-time covariance matrix (bottom). Each covariance block $C_i$ is the covariance for a single lag, with shape $n_p \times n_p$, and together they create a single space-time covariance matrix $C$ with size $n_sn_p \times n_sn_p$. The covariance matrix takes this form because the 2d images are flattened into 1d vectors, which are then joined to make an $n_p \times n_s$ 1d vector, which is multiplied by its transpose to create this matrix.}
    \label{fig:stKLIP-cartoon}
\end{figure*}

It is then necessary to create a block diagonal covariance matrix of size $n_s \times n_{\rm p}$ by $n_s \times n_{\rm p}$, as illustrated in Figure \ref{fig:stKLIP-cartoon}, from the mean-subtracted image sequence. Each block is the covariance at a given time lag, with the block diagonal as lag zero (spatial covariance). If only lag zero is used, the mathematics here reduces down to baseline (spatial) KLIP, as described in Section \ref{subsec:KLIP}. Lags should be chosen based on the translation time of the smallest relevant feature within the field of view at the focal plane up to the full crossing time of the wind across the telescope aperture. This is an additional tuneable parameter to consider when optimizing the algorithm, in addition to the number of modes.

The following computations mirror baseline KLIP, but, in practice, are potentially more computationally expensive due to the larger size of the covariance matrix used in the eigendecomposition. The steps of stKLIP are as follows:
\begin{enumerate}
    \item Subtract the mean image over the whole reference set, then partition the reference set into image sequences. These image sequences have length $n_s = n_l = 2L+1$ where $L$ is the largest number of timesteps (lags) away from the central image and $n_l$ is the total number of timesteps (lags) in the sequence.  (The following steps will be repeated over each image sequence, such that every image, with the exception of $L$ images at each end, is at some point the central image. Therefore, for $n_{i}$ images, there will be $n_{i}-2L$ image residuals at the end of this process.)
    
    Similarly to KLIP, the reference set/target image set $S$ (which in this implementation are the same) are unraveled into one-dimensional vectors $\vec{s}$ of length $n_s \times n_p$, as seen below.
    \begin{eqnarray}\label{stklipmatrix}
       S = \begin{bmatrix}
S_{1,1} & S_{1,2} & \hdots & S_{1,n_p}\\
S_{2,1} & S_{2,2} & \hdots & S_{2,n_p}\\
\vdots & \vdots & \ddots & \vdots\\
S_{n_s,1} & S_{n_s,2} & \hdots & S_{n_s,n_p}
\end{bmatrix}
\\
\vec{s} =\begin{bmatrix}
S_{1,1} & \\
S_{1,2} & \\ 
\vdots & \\
S_{1,n_p} & \\
\vdots &\\
S_{n_s,n_p}
\end{bmatrix}
    \end{eqnarray}
    \item Compute the $[n_sn_p, n_sn_p]$ size covariance matrix $C$ of the image sequence\rev{s}. In practice, this is more straightforward when done by computing the covariance of each image pair ($C_i$) and then arranging them in the block diagonal ordering shown in Figure \ref{fig:stKLIP-cartoon}.
    \item Perform an eigendecomposition on the covariance matrix, obtaining $n_sn_p$ eigenvalues ($\vec{\lambda}$) and a matrix eigenvectors ($V$) of size $[n_sn_p,n_sn_p]$ containing individual eigenvectors $\vec{v}$.
    \begin{equation}\label{eigh2}
    C\vec{v}_j = \lambda_j\vec{v}_j
\end{equation}
\begin{equation}
    \lambda_1 > \lambda_2 > \lambda_3 > \hdots \lambda_{p}
\end{equation}
    \item Choose a number of modes $n_m$, reducing the vector of eigenvalues and matrix of eigenvectors to sizes $n_m$ and $[n_m,n_sn_p]$ respectively. The matrix of eigenvectors contains $n_m$ rows of eigenvectors each with length $n_sn_p$, such that $V_{i,j} = ({\vec{v}_j})_i$. 
    \begin{equation}
    V = \begin{bmatrix}
V_{1,1} & V_{1,2} & \hdots & V_{1,n_sn_p}\\
V_{2,1} & V_{1,1} & \hdots & V_{2,n_sn_p}\\
\vdots & \vdots & \ddots & \vdots\\
V_{m,1} & V_{m,2} & \hdots & V_{n_m,n_sn_p}
\end{bmatrix}
\end{equation}
    \item Obtain image coefficients through inner product of chosen eigenvectors and image sequence, similar to Equation \ref{eq:inner}.
    \begin{equation}\label{eq:inner2}
    \vec{q} = V \cdot \vec{s} = \begin{bmatrix}
q_{1}\\
q_{2} \\
\vdots \\
q_{n_m}
\end{bmatrix}
\end{equation}
    \item Project the image sequence back into pixel space to obtain a reconstructed sequence $\vec{\hat{s}}$ with central image $\hat{\vec{\psi}}_{k}$, again mirroring Equation \ref{eq:proj}. Note: For ease of implementation, we have calculated the entire sequence, but projecting only onto the central image may improve efficiency.

  \begin{equation}\label{eq:proj2}
  \vec{\hat{s}} = \vec{\hat{q}}^T \cdot 
  {V}
  \end{equation}
  \begin{equation}
  \hat{\vec{\psi}}_{k} = [\vec{\hat{s}}_{n_p((n_l+1)/2 - 1)}\hdots \vec{\hat{s}}_{n_p(n_l+1)/2}]
\end{equation}
    \item Perform PSF subtraction using the central image.
    \begin{equation}
        \epsilon \vec{a}_{k} = \vec{s}_{k} - \hat{\vec{\psi}}_{k}
    \end{equation}
    \item Iterate through the above steps such that each image is the central image of a sequence of length $n_s$, resulting in a set of residuals $\epsilon \vec{a}_{k,j} = [\epsilon_0 \vec{a}_{k,0}, \epsilon_1 \vec{a}_{k,1}, \hdots, \epsilon_{n_s} \vec{a}_{k,n_s}]$.
    \item Compute mean of image sequence residuals to output an averaged residual, $\vec{r}_{k,{\rm avg}}$.
    \begin{equation}
    \vec{r}_{k,{\rm avg}} = \frac{1}{n_s}\sum_{j=0}^{n_s} \epsilon_j \vec{a}_{k,j}
\end{equation}
\end{enumerate}

Once our image sequence is projected into the new subspace in Step 6, we have two options for PSF subtraction: subtract the residuals from the whole sequence used, or subtract only from the central ``target'' image. We use a central target image to take advantage of speckle motions in timesteps both before and after. We then iterate through the full data set, as described in Step 8, performing stKLIP and PSF subtraction, so that each image is the central image of some image sequence with length $n_s=n_l=2L+1$. This outputs a sequence of image residuals that is of length $n_{i}-2L$. In Step 9, we then average over the number of timesteps to output an averaged residual.

There are possibilities for improving the algorithm, such as by exploiting the symmetry in the covariance matrix $C$ in order to hasten the process of updating the eigenimages; however, we leave this for future work. Further improvements are discussed in Section \ref{sec:discussion}.


\section{Algorithm Development} \label{sec:methods}

In Section \ref{subsec:sttheory}, we showed that we expect non-zero space-time covariance to exist in speckle noise. In Sections \ref{subsec:KLIP} and \ref{subsec:stKLIP}, we showed the mathematical framework for an algorithm to exploit these statistics for image processing and PSF subtraction. 

In this section, we illustrate aberrations of increasing complexity to examine their covariance structure and test the application of stKLIP. These tests and simulations are described in \ref{subsec:analytic}, for initial proof of concept. Section \ref{subsec:MEDIS} describes the algorithm application to simulated data and calculations of possible contrast gains in the algorithm's current form; here we also discuss selection criteria for the choices of number of modes and lags. Analyzing these data sets also requires some computational optimization, which is described in \ref{subsec:iterative}. In the following Section \ref{sec:results}, we will discuss the results of these applications of stKLIP.

\subsection{Foundational Tests} \label{subsec:analytic}

Our first step was to create and implement simple test cases in one and two dimensions to demonstrate that our theoretical expectations from Section \ref{subsec:sttheory} are valid and ensure that our algorithm reduced image variance as expected. A one-dimensional case allows us to directly compare a simulated covariance matrix with one calculated from the analytic theory in Section \ref{subsec:sttheory}, serving as a test of the relationship between pupil plane covariance and focal plane covariance. Then, a two-dimensional case serves as a first in implementing the algorithm, ensuring that the algorithm reduces variance on a well-understood simple case before moving onto more complex atmospheric simulations.

\subsubsection{One-Dimensional Test of Pupil/Focal Covariance Relationship}
To begin, we created a simple one-dimensional model of two interfering speckle PSFs, which are simply two sinusoids  with slightly different frequencies in the pupil plane. We first use this simple sinusoidal model to compare the simulated space-time covariance to the predicted behavior from theory, to show how a set of input aberrations in the pupil plane corresponds with the resulting focal-plane space-time covariance. Although the algorithm does not require pupil plane covariances, this test is done to further establish the existence of the focal plane covariances that we seek to harness.

To create the 1-d speckle model, first we must create a grid setup for evaluating the wavefront in the pupil and focal planes. These are parameterized in units of $D/\lambda$ and $\lambda/D$ respectively, where $\lambda$ is our wavelength of observation, assuming monochromatic light. 
Keeping these units preserves the Fourier duality relationship, and they can be converted to more conventional units if the focal length is known.

\begin{figure}
    \centering
    \includegraphics[width=1\linewidth]{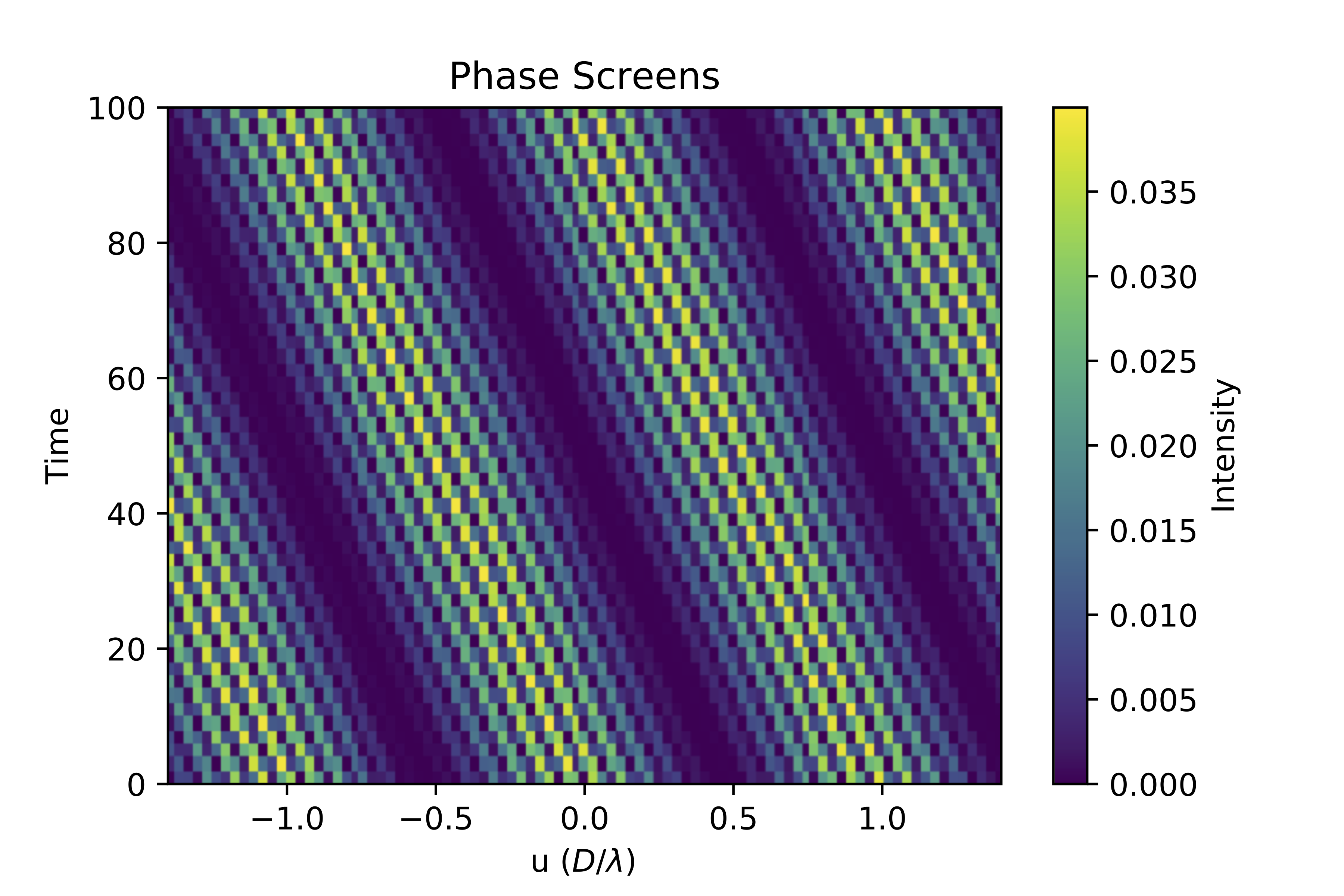}
    \includegraphics[width=1\linewidth]{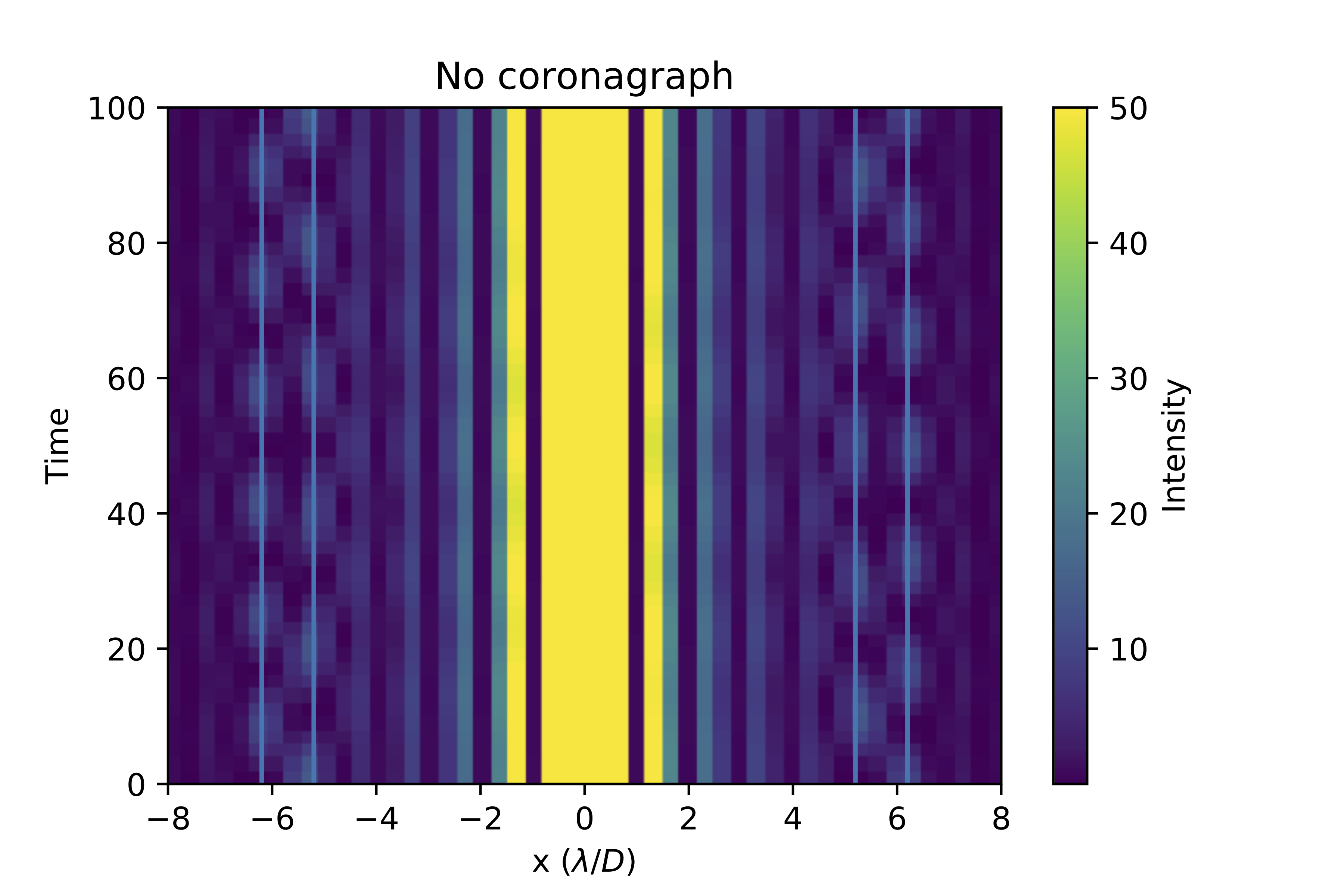}
    \includegraphics[width=1\linewidth]{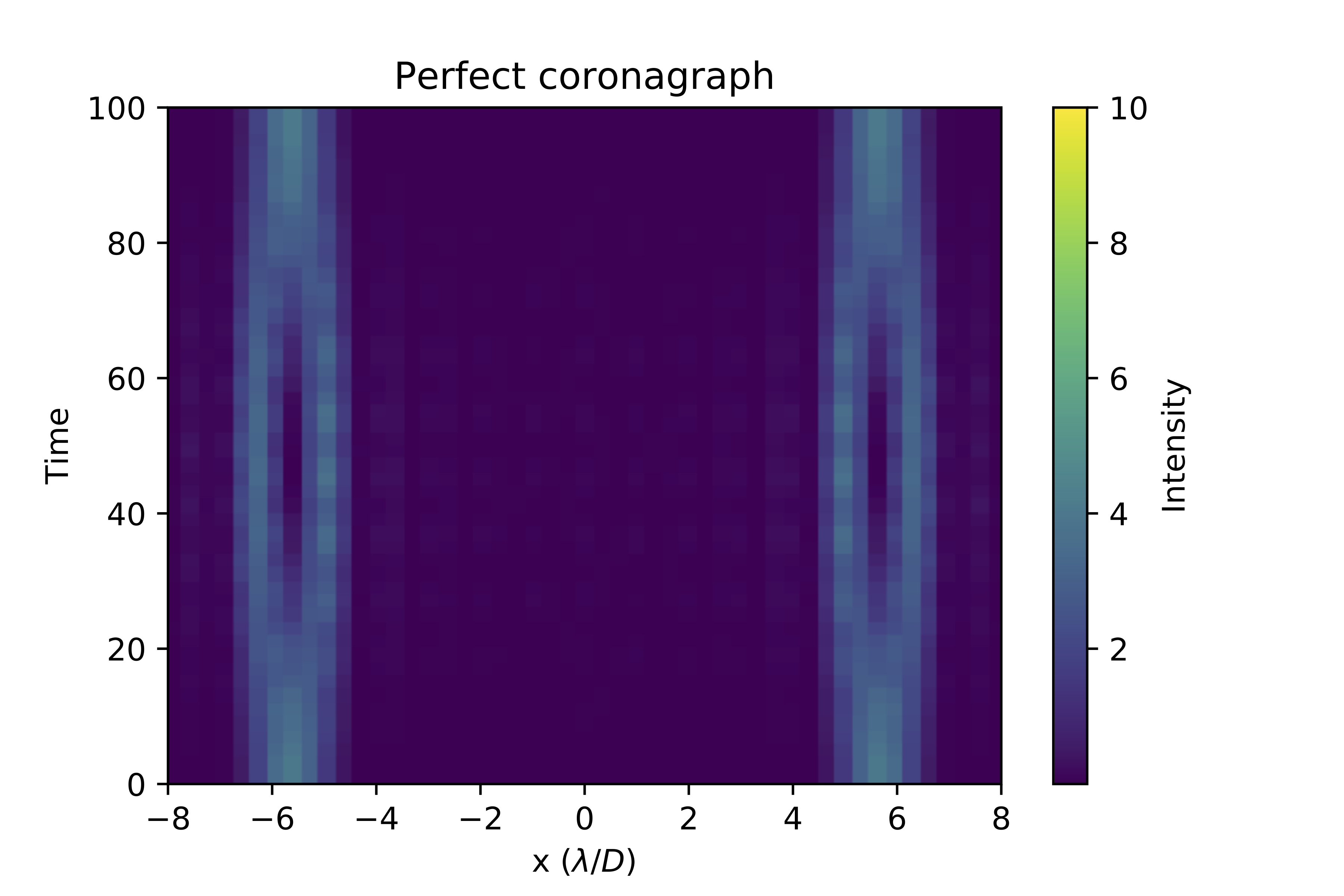}
    \caption{One-dimensional demonstration of speckle interference. Two sinusoidal perturbations in the pupil plane interfere to create moving speckles in the image plane. Top: 1d phase screen with interfering sinusoids over time. Middle: 1-d intensity over time without a coronagraph, showing the Airy pattern. Bottom: 1-d intensity over time with a coronagraph, with the speckles' relative evolution appearing more clearly due to the lack of coherent light, $C(\vec{x})$. This simulation is used as a test of the space-time speckle covariance theory in Section \ref{subsec:sttheory}.}
    \label{fig:1d-test}
\end{figure}

The next critical piece is to define the entrance aperture in the pupil plane. This pupil function sets the amplitude $A$ of the electric field ($E = Ae^{i\phi}$), and is simply a top-hat function ($\Pi(u)$, 1 inside a given region and 0 outside). We also apply a translating phase screen (shown in the top panel of Figure \ref{fig:1d-test}) to the pupil, which is where phase aberrations are accounted for. We use a simple perturbation of two superimposed sinusoids with similar periods/frequencies, so that the wings of their PSFs overlap. This set-up is like simulating one layer of frozen flow translating across the telescope's aperture. These perturbations are small ($\ll$ 1 radian), consistent with the high-contrast regime. 

We then perform the necessary Fourier transform to retrieve the focal-plane electric field. By doing this for the pupil function with no perturbations, we retrieve what we would see in an ideal case for a uniformly illuminated pupil; this is also what would be blocked if we had a perfect coronagraph. We subtract this ``perfect'' case from the case with the sinusoidal perturbation, performing the action of the coronagraph and suppressing light from the unaberrated portion of the wavefront.

\begin{figure*}
    \centering
    \includegraphics[width=0.6\linewidth]{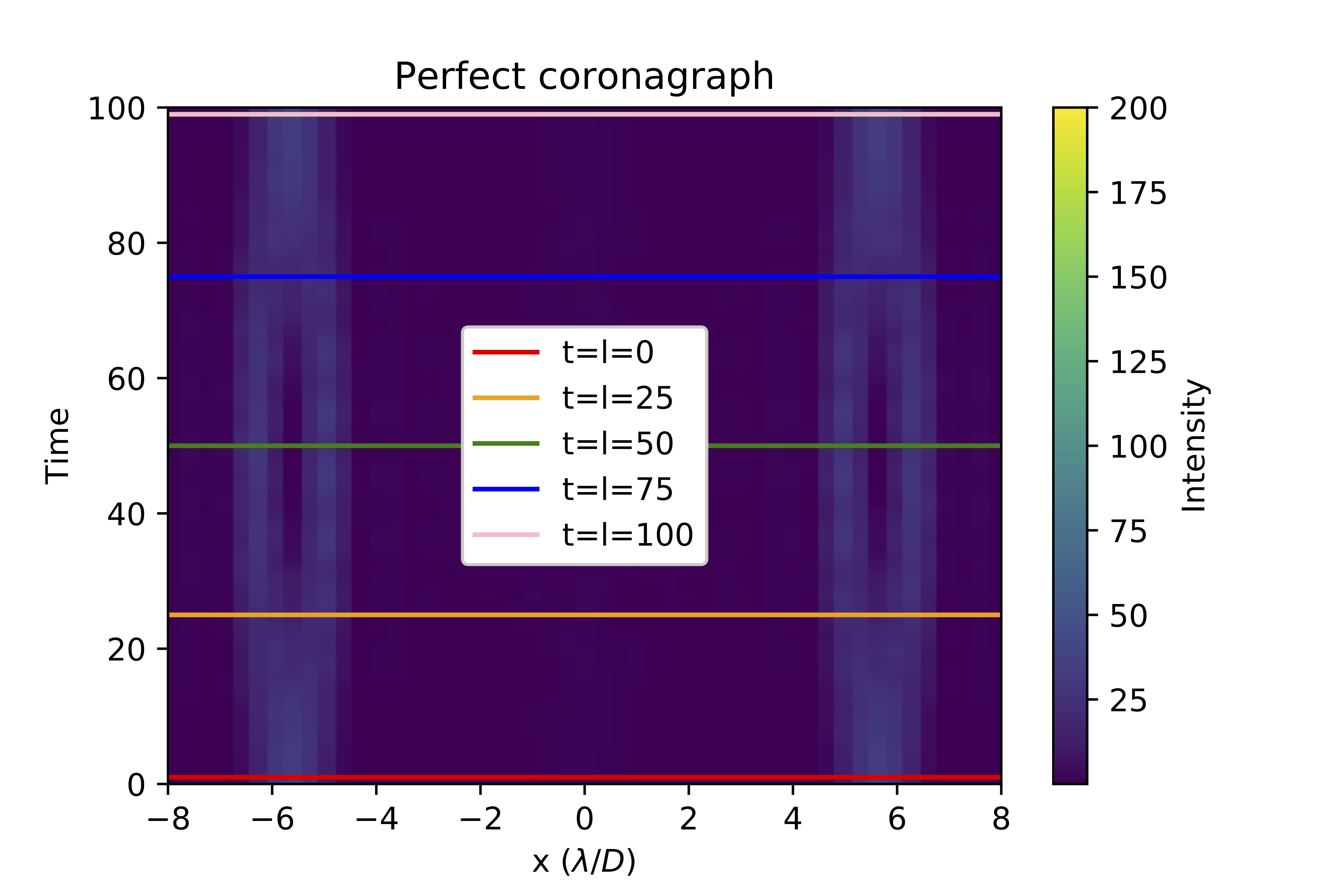}
    \includegraphics[width=0.9\linewidth]{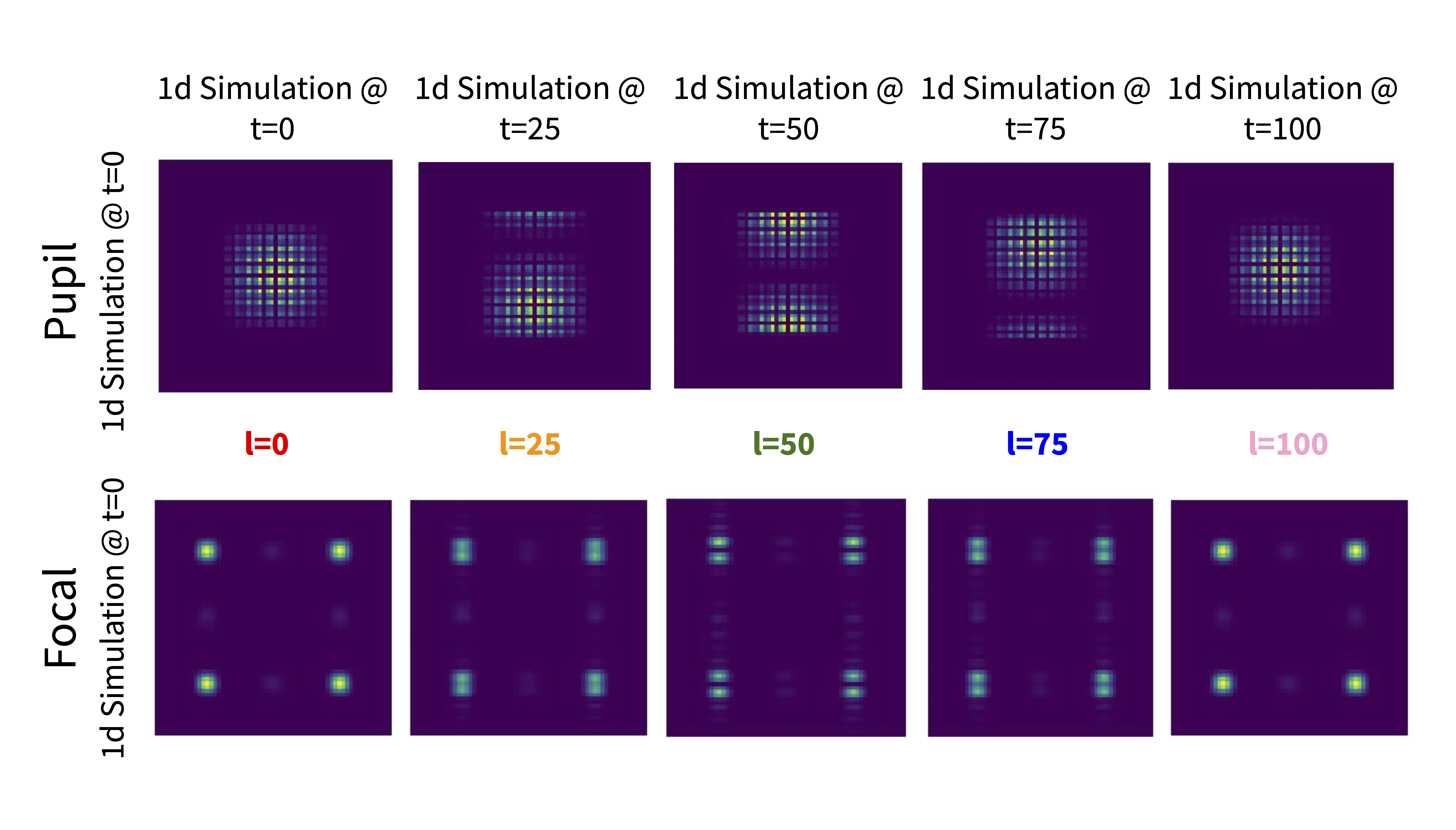}
    \caption{Space-time covariance matrices for pupil plane (middle) and focal plane (bottom) of a 1-d model of two sinusoids with different frequencies -- as illustrated in the top panel of Figure \ref{fig:1d-test} -- with an annotated view of the simulation (top). These matrices show a symmetric pattern that changes with the number of lags used, due to the change in the speckles' relative locations. At lags 0 and 100, the peaks are due to the alignment of the speckles' peaks, as marked in the top panel; lag 25 illustrates the lower covariance when the speckles are in slightly different places, and lag 50 shows two lower intensity peaks when the speckles are separated. Importantly, for a given non-zero lag, there are non-zero terms, indicating that there are temporal correlations.}
    \label{fig:cov1d}
\end{figure*}

A one-dimensional case (Figure \ref{fig:1d-test}) illustrates the relative evolution of two neighboring speckles created from atmospheric perturbations. Atmospheric theory (as in Section \ref{subsec:sttheory}), in particular the frozen flow assumption, predicts a symmetrical space-time covariance structure, which can be computed for a 1-d model with a top-hat pupil function ($\Pi(u)$), two sinusoidal functions in the pupil, and no uniform illumination in the pupil ($C(x)=0$). We carried out these calculations in two ways. First, we solved the integrals in Section \ref{subsec:sttheory} for the simple two sinusoid situation using Fast Fourier Transforms (FFTs). Second, we began with an array describing the sinusoidal ``phase screen'' and simulated propagation through an optical system using FFTs. 

The variation in pupil and focal plane covariance over various time lags, as shown in Figure \ref{fig:cov1d}, can be clearly interpreted based on the locations of the two interfering speckles. These matrices show a symmetric pattern that changes with the number of lags used, due to the change in the speckles' relative locations. At lags 0 and 100, the peaks are due to the alignment of the speckles' peaks, as marked in the top panel; lag 25 illustrates the lower covariance when the speckles are in slightly different places, and lag 50 shows two lower intensity peaks when the speckles are separated. Importantly, for a given non-zero lag, there are non-zero terms in both the pupil and focal plane covariances, indicating that there are temporal correlations.

This simulation further demonstrates the claim that a simplified frozen flow scenario in the pupil can create calculable space-time covariances in the focal plane, and validates our use of this simple test case to test stKLIP.

\subsubsection{Two-Dimensional Test Case for Algorithm Development}

\begin{figure}
    \centering
    \includegraphics[width=1\linewidth]{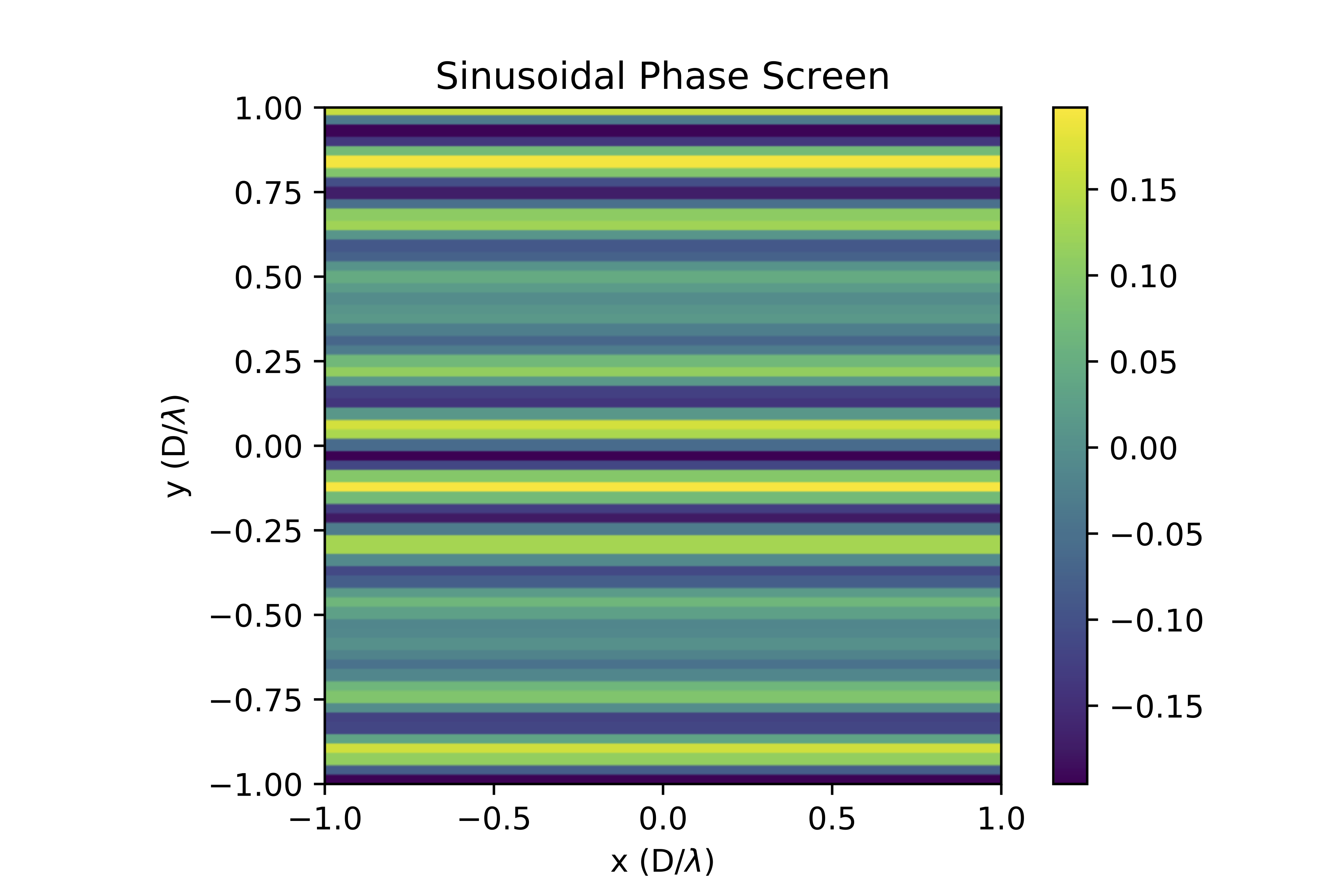}
    \includegraphics[width=1\linewidth]{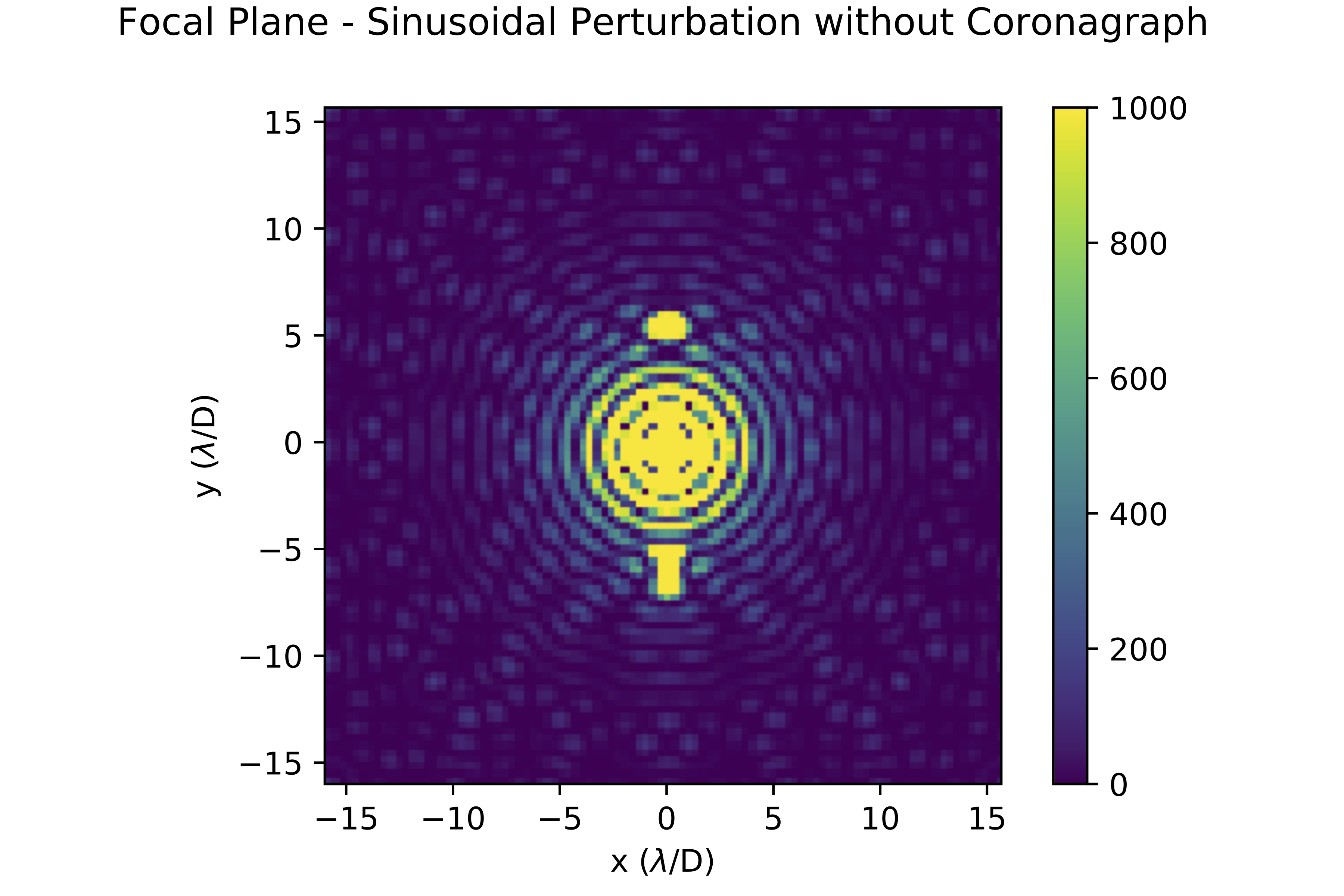}
    \includegraphics[width=1\linewidth]{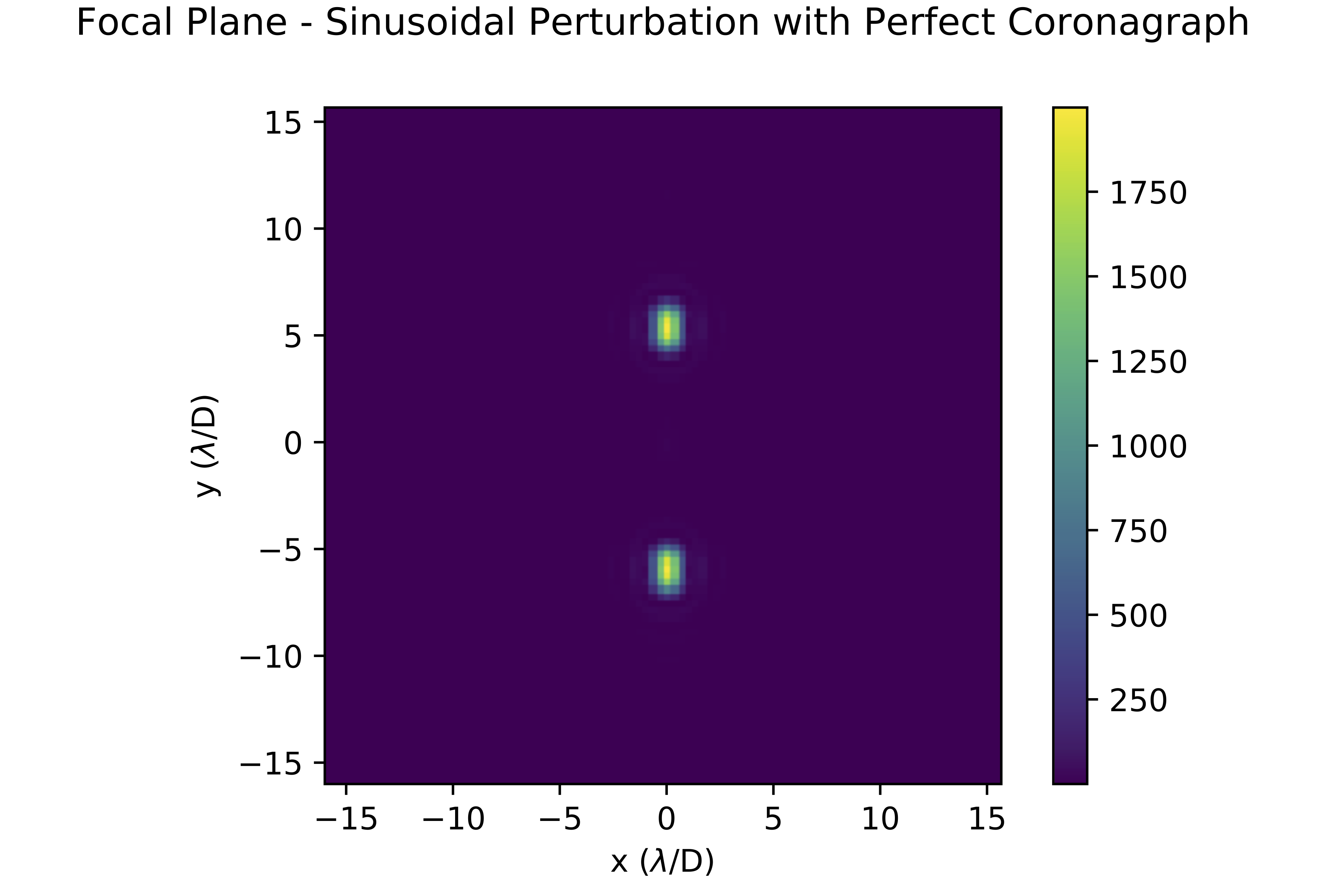}
    \caption{Two-dimensional test of speckle interference. A sinusoidal phase screen (top) produces a speckle pattern imposed on an Airy disk (middle). Subtracting the PSF of a model without perturbations, we simulate observations of this sinusoidal perturbation with a ``perfect'' coronagraph (bottom). All images depict the intensity ($I = |E|^2$). This simulation is used as a troubleshooting step for a first implementation of the stKLIP algorithm.}
    \label{fig:2d-test}
\end{figure}

In order to ensure that the algorithm is behaving according to our expectations -- that it will reduce the image variance -- we expand this one-dimensional test case into two-dimensions to make an image sequence of the two time-varying, interfering speckles. We use this idealized test case as a check against our expectations for our stKLIP implementation, and for a first test of efficacy, comparing the reduction in image variance between three data processing methods: mean-subtraction, KLIP, and stKLIP. The setup is the same as the above one-dimensional test case, but in two dimensions, with a circular aperture instead of a top hat as the pupil function. We create a series of images at various time steps as the input to stKLIP, shown in Figure \ref{fig:2d-test}.

Although there are two tuneable parameters for stKLIP --- number of modes (e.g. number of eigenimages used in the projection) and number of lags, as described in Sections \ref{subsec:KLIP} and \ref{subsec:stKLIP} --- we only test one set of modes and lags (10 modes, 2 lags) with this simple test case and leave further exploration of these parameters for later testing (see Section \ref{subsec:MEDIS}). We similarly use 10 modes for KLIP to make the comparison fair.

\begin{figure*}
    \centering
    \includegraphics[width=\linewidth]{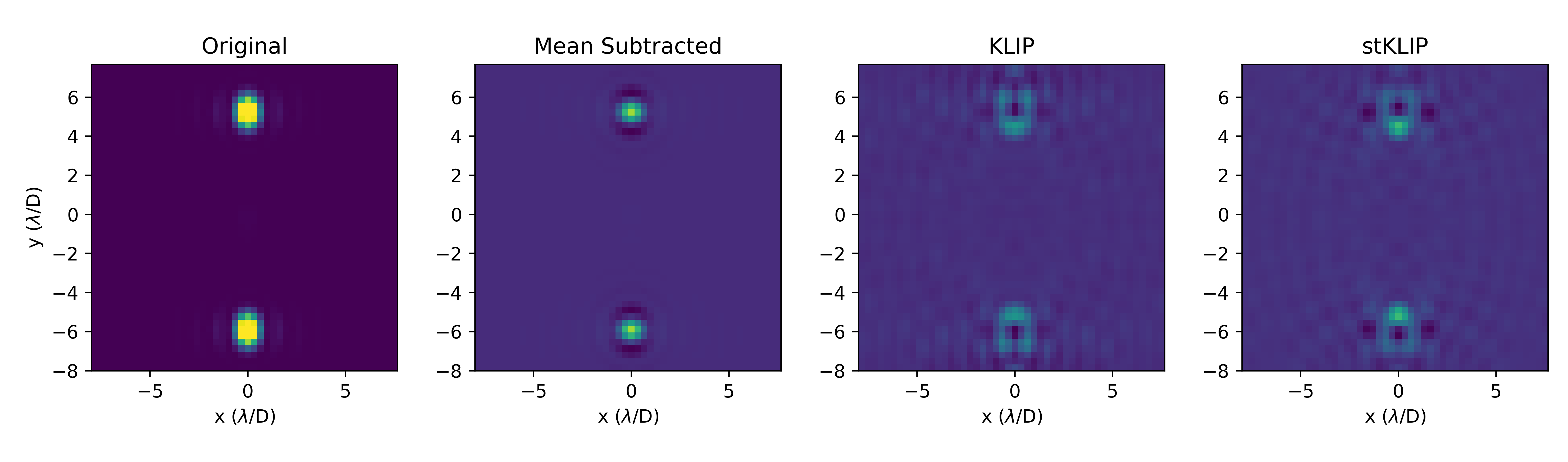}
    \caption{One frame of the input sequence (left) for the simple two-sinusoid test case with a coronagraph, with the residuals after PSF subtraction using mean-subtraction, KLIP, and stKLIP, showing a clear reduction in speckle intensity. Both stKLIP and baseline KLIP reduce image variance by a factor of at least 5.7 from the original image, an improvement over simple interventions like mean-subtraction. Although stKLIP does not improve upon KLIP in this limited test case, it is important to remember that we have not optimized for modes and lags in this scenario; this step was intended for troubleshooting, not rigorous characterization of the algorithm.}
    \label{fig:stklip-basic-test}
\end{figure*}

In this simple test case, KLIP and stKLIP reduce the variation in the image by factors of 6.8 and 5.7, respectively. Although stKLIP does not improve upon KLIP in this limited test case, it is important to remember that we have not optimized for modes and lags in this scenario; determination of performance is left for more rigorous and realistic tests in the following section, \ref{subsec:MEDIS}. They both outperform simple interventions, such as subtracting the mean of the image, in reducing the total variation in the image, as shown in Figure \ref{fig:stklip-basic-test}. To summarize, this 2d test was performed to demonstrate that the overall image variance decreases after projecting out modes of variation with stKLIP, as qualitatively expected, and in that sense the test can be considered successful.


\subsection{Simulated AO Residual Tests} \label{subsec:MEDIS}

We then wanted to test stKLIP on a more realistic atmospheric phase screen and again measure potential contrast gains. To this end, we created a set of simulated observations to represent AO residuals and performed stKLIP on them for a variety of different modes and lags. We measure contrast curves and companion SNR for four methods of post-processing in order to understand the effectiveness of our new method: stKLIP, baseline/spatial KLIP, mean-subtraction, and no post-processing. Results from these tests are described in Section \ref{sec:results} and discussed further in Section \ref{sec:discussion}. In this section, we first detail the methods used to create the simulated data set, then the methods for computing contrast curves and SNR on the processed data.

To create the simulated data set, we use a simulator specifically designed for high-contrast imaging with next-generation detectors, such as MKIDs, called \texttt{MEDIS} (the MKID Exoplanet Direct Imaging Simulator), the first end-to-end simulator for high contrast imaging instruments with photon counting detectors \citep{dodkins2018development,dodkins2020medis}.

\begin{figure}
    \centering
    \includegraphics[width=0.9\linewidth]{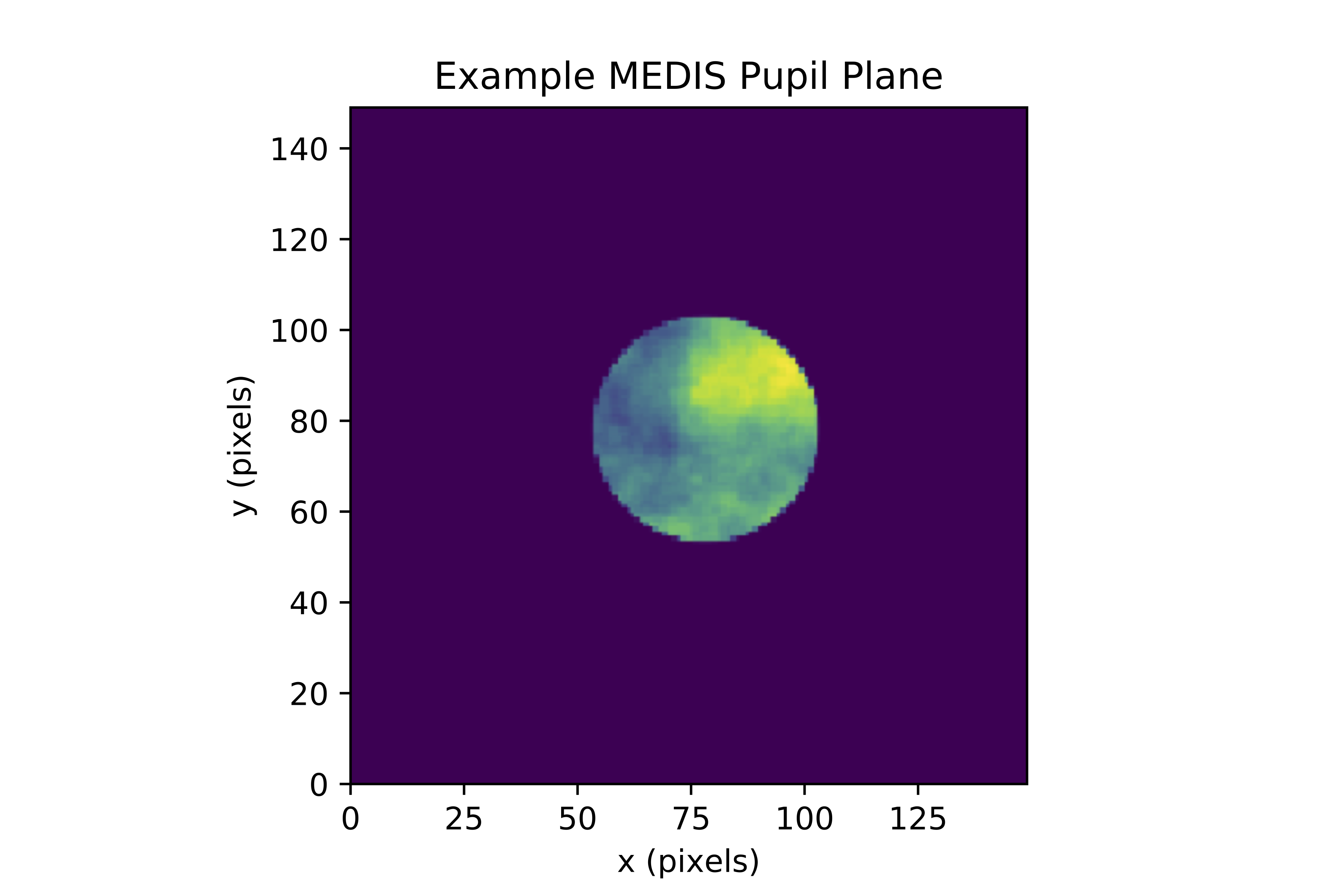}
    \includegraphics[width=0.9\linewidth]{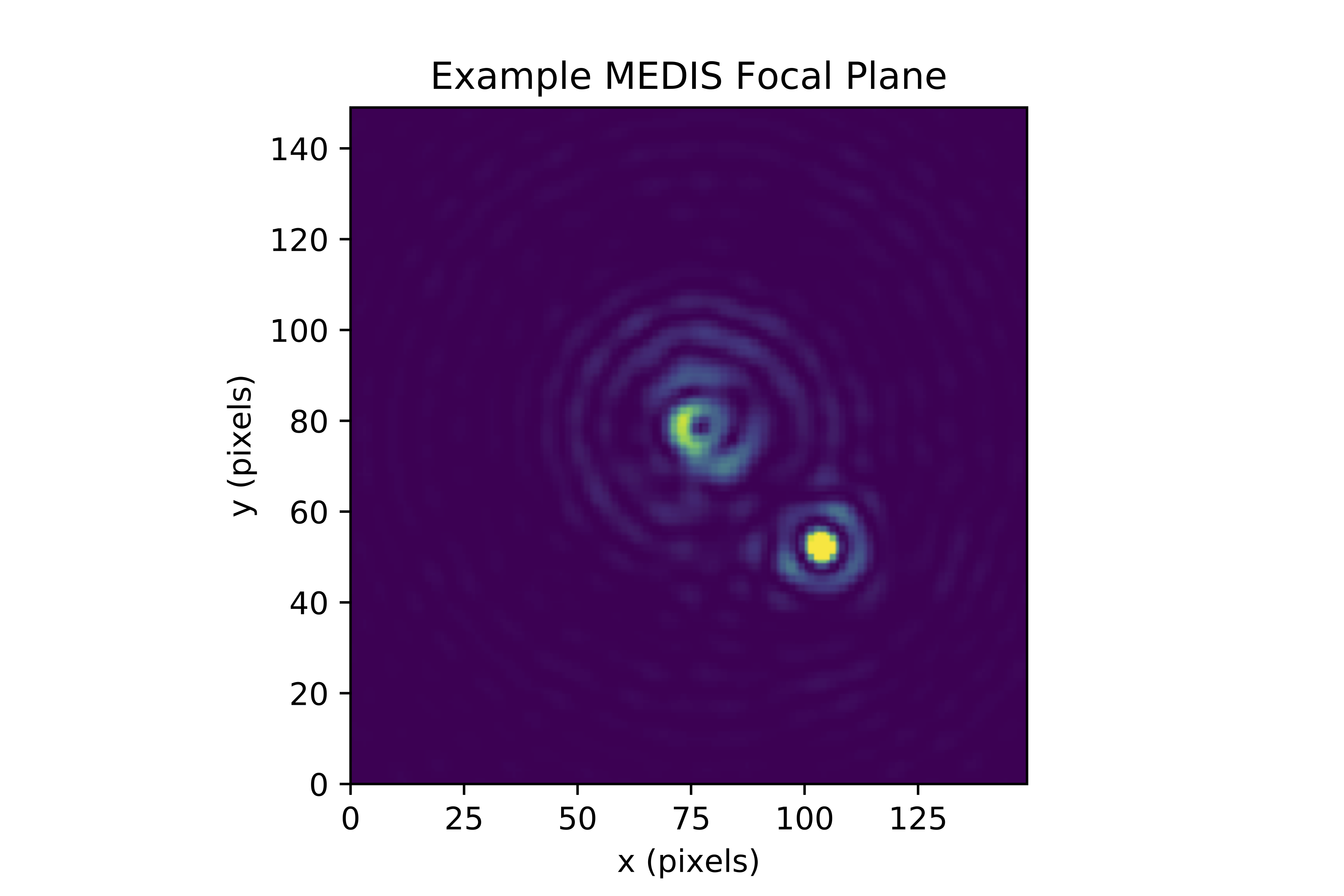}
    \caption{Examples of MEDIS simulations. (Top) Pupil plane, illustrating the phase screen. (Bottom) Focal plane, with a clearly bright companion object. These simulations are used as a preliminary test of stKLIP's efficacy and potential; however, there is a large parameter space to explore beyond the scope of this work.}
    \label{fig:MEDIS-sim}
\end{figure}

\texttt{MEDIS} generates atmospheric phase screens with \texttt{HCIPy} \citep{por2018hcipy}. These phase screens use models of Kolmogorov turbulence, and we use the simplest option of a single frozen flow layer. Then, \texttt{MEDIS} uses \texttt{PROPER} to propagate the light through the telescope under Fresnel diffraction, including both near- and far-field diffraction effects \citep{krist2007proper}. Separate wavefronts are propagated for each object in the field --- the host star, and any companion planets. \texttt{MEDIS} also includes options to introduce coronagraph optics, aberrations (like non-common path errors), and realistic detectors. \texttt{MEDIS} outputs the electric field or intensity at specified locations in the optical chain, such as the pupil and focal planes in our case, as shown in Figure \ref{fig:MEDIS-sim}. 

Given the wide range of parameters available in \texttt{MEDIS}, we had to make decisions on what to use for the \texttt{MEDIS} simulations used to test stKLIP. For these simulations, we implement a telescope with 10 meter diameter, similar to the Keck Telescopes. We begin with a case without adaptive optics for simplicity. For this, the sampling rate needs to be a few milliseconds, a few times oversampled compared to the smallest temporally resolvable features given the field-of-view (FOV) under consideration. The number of frames is chosen to create a total observation time of 30 seconds (6,000 frames at 0.005 second sampling) 
to recreate a realistic observation and attain a sufficient number of independent samples. The grid size is significantly larger than the area of interest (256 $\times$ 256 pixels) to avoid edge effects. However, we choose a region size / FOV that is significantly smaller than our whole grid (100 $\times$ 100 pixels) to make this problem more computationally tractable.

The simulation includes atmospheric parameters, such as the Fried Parameter ($r_0$), a length scale for coherence in the atmosphere, and the structure constant ($C_n$), \rev{a description of turbulence strength over multiple atmospheric layers.} \rev{The atmospheric model we use is a simple single layer of extremely mild Kolmogorov turbulence, with} $r_0 > 10$ \rev{m,} since we want $r_0 \gg D$ to stay in the high-contrast regime of small phase errors. \rev{Note: this simulated atmosphere is not realistic in ground-based imaging, but we chose these parameters to approximate the high-contrast regime without simulating adaptive optics and introducing additional parameters. While our numerical experiments will depend on the input power spectrum, our primary aim was to assess the characteristics of a second-order statistical analysis of the linearized system} (Equation \ref{eq:approx})\rev{, rather than impacts of the particulars of the wavefront error power spectrum.} It is worth exploring how different atmospheric conditions (e.g. a \rev{smaller} $r_0$ value) would change the effectiveness of this method, but that is beyond the scope of this initial investigation.

We choose a vortex coronagraph \citep{mawet2009vector}, since it is the closest to an ``ideal'' coronagraph of the options available in \texttt{MEDIS} (e.g. closest to perfect cancellation of the spatially coherent wave), thanks to its small inner working angle \citep{guyon2006theoretical}. We want an ideal detector since, for this initial investigation, we are not yet interested in how detector noise/error affects this method. We also include one companion object that would be readily detectable given current capabilities (a contrast of $5 \times 10^3$), in order to enable SNR measurements of an injected companion for various post-processing methods including stKLIP. As mentioned in Section \ref{subsec:sttheory}, lags should be chosen based on crossing times and relevant features. In these simulations, this ranges from 2 to 10 timesteps (0.01 to 0.05 seconds) for a wind speed of 5 m/s and 5 millisecond sampling. Future work should test a further range of lags, up to 400 timesteps (2 seconds, or one full crossing time), but our current method is computationally limited as mentioned in Section \ref{subsec:iterative}. In this investigation, we also test a range of modes from 1 to 500.

Although these simulations are computationally expensive, \texttt{MEDIS} is capable of parallel processing, except in cases where AO parameters require serialization. We take advantage of this capability by using UCLA's Hoffman2 Cluster. The resultant data sets are quite large, and require inventive ways of computing the necessary statistics without loading the full array into memory, described further in Section \ref{subsec:iterative}. These simulations show us how realistic space-time covariance differs from the idealized case, and allow us to begin to test the effectiveness of our new method. 

Metrics of efficacy used in this study are measurements of variance, signal, noise, signal-to-noise ratios (SNR), and contrast curves. Variance is simply computed over the whole 100$\times$100 pixel residual image using \texttt{numpy.var}. Signal is computed using aperture photometry (via \texttt{photutils}), centered on the simulated companion. Noise is similarly computed using aperture photometry by taking the standard deviation of a series of apertures in an annulus at the same separation as the simulated companion. SNR is then the ratio of these two measurements. Contrast curves are estimated using aperture photometry at various distances from the image center and dividing by the aperture photometry measurement of the unmasked (e.g. no coronagraph) peak, then adjusting by the signal throughput; the throughput here is estimated as the signal after processing divided by the signal before data processing. These various metrics are computed for the original images, as well as different post-processing scenarios, to understand the relative efficacy of stKLIP. Results are described in Section \ref{sec:results}.

\subsection{Iterative Statistics Calculations} \label{subsec:iterative}

There are two key computational challenges for large data sets such as those produced by \texttt{MEDIS}: memory access and computational complexity. Simulations with \texttt{MEDIS} for a realistic observing sequence based on our criteria above can be on the order of 100GB, which can pose challenges to RAM-based manipulation for the calculation of mean and covariance given our current computing resources. To address this problem, we implemented the framework for iterative statistics calculations set forth in \citet{savransky2015sequential}.

In order to perform a KLIP-style calculation, we first need to compute second-order statistical quantities for a data set of $n$ samples $\vec{x}_i$, such as the mean and covariance. The formula for the calculating mean is:
\begin{equation}
\vec{\mu} \equiv \frac{1}{n}\sum_{i=1}^{n}\vec{x}_i
\end{equation}
When the mean $\mu$ is estimated from the data, the sample covariance can be calculated as follows:
\begin{equation}
C \equiv \frac{1}{n-1}\sum_{i=1}^{n}(\vec{x}_i - \vec{\mu})(\vec{x}_i - \vec{\mu})^T.
\end{equation}

These sums can be broken up into smaller iterative steps $k$, to make the calculation less memory intensive. For each step $k$, the mean can be updated with the formula
\begin{equation}
\vec{\mu}_k = \frac{(k-1)\vec{\mu}_{k-1}+\vec{x}_k}{k}
\end{equation}
and the covariance can be updated by
\begin{equation}\label{iter-spatial-cov}
S_k = \frac{k-2}{k-1}S_{k-1} + \frac{k}{(k-1)^2}(\vec{x}_k-\vec{\mu}_k)(\vec{x}_k-\vec{\mu}_k)^T.
\end{equation}

However, Equation \eqref{iter-spatial-cov} is only applicable to the spatial covariance, e.g. a time lag of zero. The space-time covariance can be calculated as 

\begin{equation}
S_l = \frac{1}{n-l-1}\sum_{i=1}^{n}(\vec{x}_i-\vec{\mu})(\vec{x}_{i-l}-\vec{\mu})^T.
\end{equation}

Following a similar protocol to \citet{savransky2015sequential}, we derived an update formula for the space-time covariance:
\begin{multline}
    S_l = \frac{1}{n-l-1} \Bigg[\sum_{i=l}^{n}\vec{x}_i\vec{x}_{i-l}^T - (n-l)\vec{\mu}\vec{\mu}^T \\
    + \vec{\mu}^T\sum_{i=1}^{l-1}\vec{x}_i +\vec{\mu}\sum_{i=n-l-1}^n\vec{x}_i^T - 2l\vec{\mu}\vec{\mu}^T \Bigg]
\end{multline}
It is identical to Equation \eqref{iter-spatial-cov}, except for the last 3 additional cross-terms. These cross-terms were directly calculated and determined to be negligibly small as the sample size becomes large relevant to the maximum lag, and thus would only be relevant in edge cases. For 1,000 samples, the error on the space-time covariance calculation is on the order of $10^{-4}\%$ or less. For 10,000 samples, the error decreases to $10^{-6}$ to $10^{-7}\%$, indicating a trend of decreasing error for an increasing number of samples. We do not plan to use fewer than 1,000 samples in a data set, so we consider this approximation to the space-time covariance acceptable and have implemented it for the tests described in Section \ref{subsec:MEDIS}.

Although the mathematics laid out in this section make covariance calculations possible, the resulting covariance matrices can be quite large, on the order of 10GB for even short test cases with small FOVs. Even with sufficient RAM for manipulation, these large covariance matrices can lead to long computation times for following steps of the algorithm. The image size and sequence length of data sets used in our stKLIP method is therefore still currently limited by memory requirements and prohibitively long execution times. This is mostly due to the eigendecomposition calculations, since the full space-time covariance matrix needs to be loaded into memory for input into \texttt{scipy.linalg.eigh}. As we proceeded with larger data sets, we chose to perform a standard eigendecomposition with \texttt{scipy.linalg.eigh} \rev{using the default backend (C LAPACK \texttt{evr})} but limited the maximum number of eigenvalues/eigenvectors computed, since many of the smaller eigenvalues only capture noise and are not necessary for this process. \rev{There may be more optimal choices for the eigendecomposition algorithm, but such optimization is left for future work.}

Another possible solution to mitigate this bottleneck would be using an iterative eigendecomposition. This could theoretically be done with the NIPALS (Nonlinear Iterative Partial Least Squares) algorithm \citep{risvik2007principal}. However, applying the NIPALS algorithm is not straightforward for this problem; our space-time covariance matrix is currently assembled from various spatial covariance matrices, and considerable changes would need to be made to NIPALS to accommodate a space-time calculation instead of a solely spatial one, since the NIPALS algorithm relies on a data matrix as input instead of a covariance matrix. Future iterations of this algorithm could also make use of the \texttt{dask} package for parallelization of computations to help speed up run time, but as of this writing an eigendecomposition function (e.g. \texttt{dask.linalg.eigh}) was not yet implemented, although the similar \texttt{dask.linalg.svd} function could possibly be used. We leave such improvements in efficiency for future work.

\section{Algorithm Performance on Simulated AO Residual Data} \label{sec:results}

We have confirmed through theory (\S\ref{subsec:sttheory}) and simulation (\S\ref{subsec:analytic}) that space-time covariances exist for speckles in a simple high-contrast imaging system in the regime of small phase errors and short exposures. In Section \ref{subsec:stKLIP}, we defined a new algorithm, similar to Karhunen-Loe\'ve Image Processing, to take advantage of space-time covariances and improve final image contrast, with the eventual goal of detecting fainter companion objects. As shown in Section \ref{subsec:MEDIS}, we have developed an initial implementation of this space-time KLIP (stKLIP) algorithm, and demonstrated it on simulated data. In this section, we present the results of those demonstrations. It is worth noting that these tests on simulated data only explore a small range of parameter space, and are not indicative of the absolute potential of using space-time covariance in data processing. Instead, we present this as a first proof-of-concept for the possibility of this new method.

\begin{figure*}
    \centering
    \includegraphics[width=\linewidth]{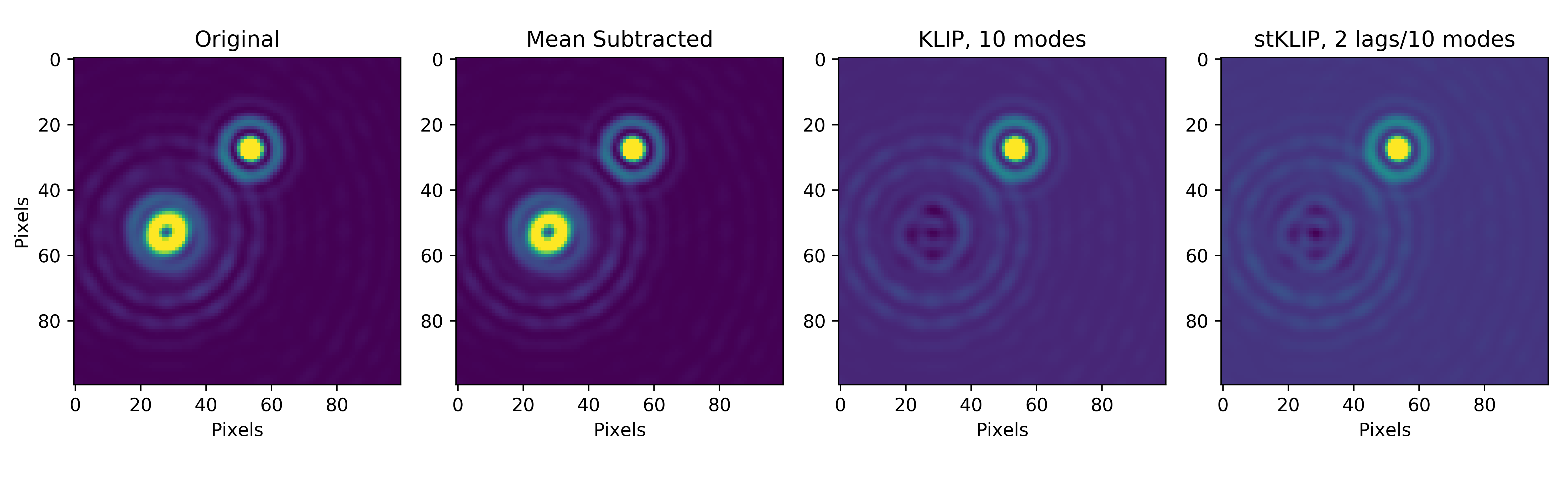}
    \caption{One frame of the input sequence (left) from \texttt{MEDIS}, with the residuals after PSF subtraction using mean-subtraction, KLIP, and stKLIP. Both stKLIP and baseline KLIP reduce image variance by a factor of $\sim$1.85 from the original image for the listed case of 10 modes and 2 timesteps lag in stKLIP.}
    \label{fig:tested-MEDIS}
\end{figure*}

An example of the images input to and output by the stKLIP processing algorithm is shown in Figure \ref{fig:tested-MEDIS}, along with a comparison to two other data processing interventions, mean-subtraction (as in Equation \ref{eq:ms}) and KLIP. For this simulated data, mean-subtraction makes such a slight improvement that in the following figures we omit it from comparison plots, as it would be almost precisely coincident with the original image's metrics.

To quantitatively measure the efficacy of our stKLIP data processing algorithm, we computed total image variance, signal-to-noise ratios, and approximate contrast curves, as described in Section \ref{sec:methods}. To further determine the utility of this algorithm and characterize its dependence on the tuneable parameters, we also investigated the relationships between these efficacy metrics, the number of KL modes used, and the number of stKLIP lags used. We leave adjustments of the residual wavefront error statistics and companion location, among other parameters, and their effects on stKLIP's efficacy for future work.

Image variance is a primary metric for subtraction efficiency. Total image variance is reduced by almost half for both spatial / baseline KLIP and stKLIP within the first 10 modes, and variance approaches 0 around 50 modes. In this test, spatial and stKLIP are similar in their variance reduction abilities, and are both improvements on mean-subtraction and the original image. Image variance drops off steeply within the first 20 modes, indicating that most of the power is removed with only a few eigenimages required in the reconstruction. Given that only a small number of modes are required to remove the majority of the variance in the image, future applications of this algorithm could exploit this fact to reduce the computational burden by only calculating the first $n$ eigenvalues/eigenimages.

\begin{figure}
    \centering
    \includegraphics[width=\linewidth]{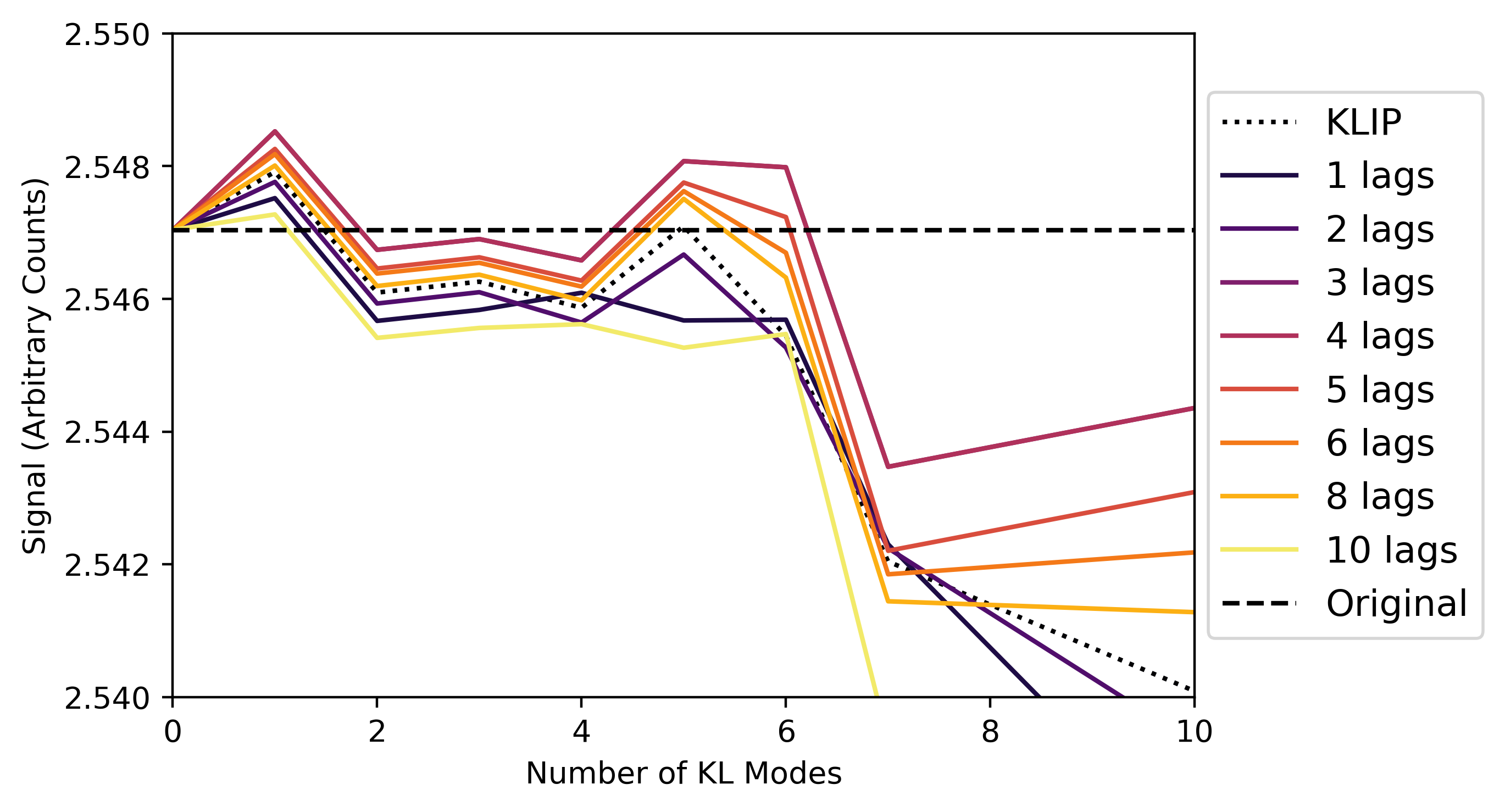}
    \caption{Companion signal over number of KL modes used in the model PSF subtraction; this figure shows that signal loss begins around 5 modes, indicating that future iterations of this algorithm would benefit heavily from implementing measures to prevent self-subtraction. Certain choices of lag (4, 5, 6, 8) show a minor improvement in signal retention over spatial (lag = 0) KLIP.}
    \label{fig:signal}
\end{figure}

For both KLIP and stKLIP on these simulated data, signal starts to be lost around 5--10 modes and drops off more steeply after $\sim$30 modes. Space-time KLIP with 4, 5, 6, or 8 lags in this scenario shows a slight edge over baseline KLIP in signal retention, as shown in Figure \ref{fig:signal}. It is worth noting that the choice of optimal number of lags depends on the wind speed and region in the image that we are most interested in. Recall from Section \ref{subsec:MEDIS} that this test uses $v = 5$\,m/s, and the companion location can be seen in Figure \ref{fig:tested-MEDIS}. Noise reduction capabilities appear very similar between KLIP and stKLIP; after about 40--50 modes, so much of the image has been removed that noise approaches zero and shows small random fluctuations, indicating that these higher modes contain less information.

\begin{figure}
    \centering
    \includegraphics[width=\linewidth]{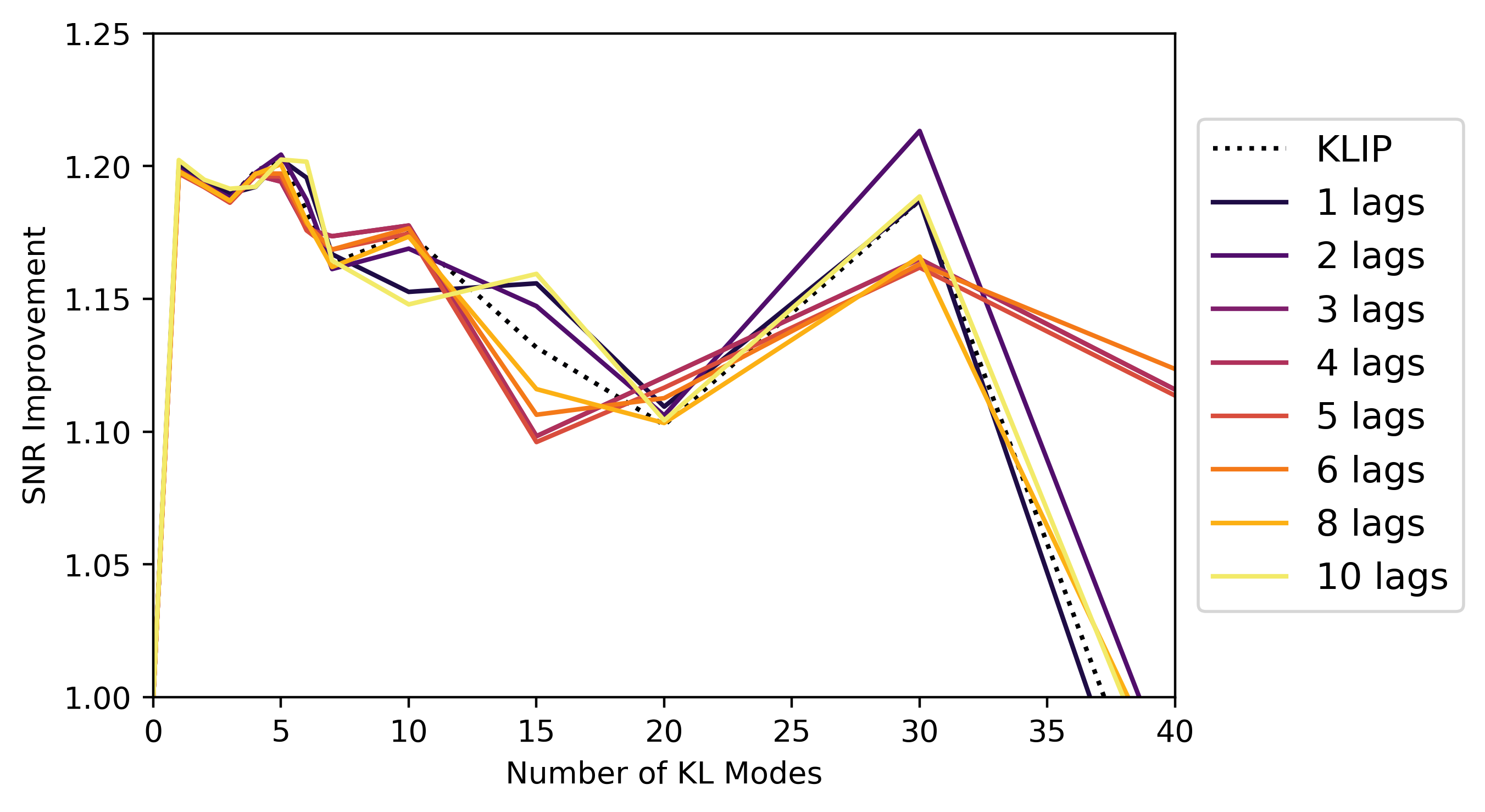}
    \caption{Companion signal-to-noise ratio (SNR) compared to the original image SNR over number of KL modes used in the model PSF subtraction; this figure shows a 10--20\% improvement over the original image using stKLIP and KLIP, with stKLIP having a slight edge (on the order of a few percent) for certain choices of lag.}
    \label{fig:SNR}
\end{figure}

Signal-to-noise ratio (SNR) shows a 10--20\% improvement over the original image within the first 40 modes, as shown in Figure \ref{fig:SNR}. The 2nd peak in Figure \ref{fig:SNR} is possibly due to small number statistics (most of the signal has been removed by then) and not a real SNR improvement. It is worth noting that the SNR shown here could improve significantly if a method is implemented to retain signal and improve throughput, which we discuss more in the following section. We again see that there is a slight advantage for certain lags over spatial (lag=0) KLIP on the order of a few percent, indicating that there is possibility for properly tuned stKLIP to outperform KLIP.

\begin{figure*}
    \centering
    \includegraphics[width=\linewidth]{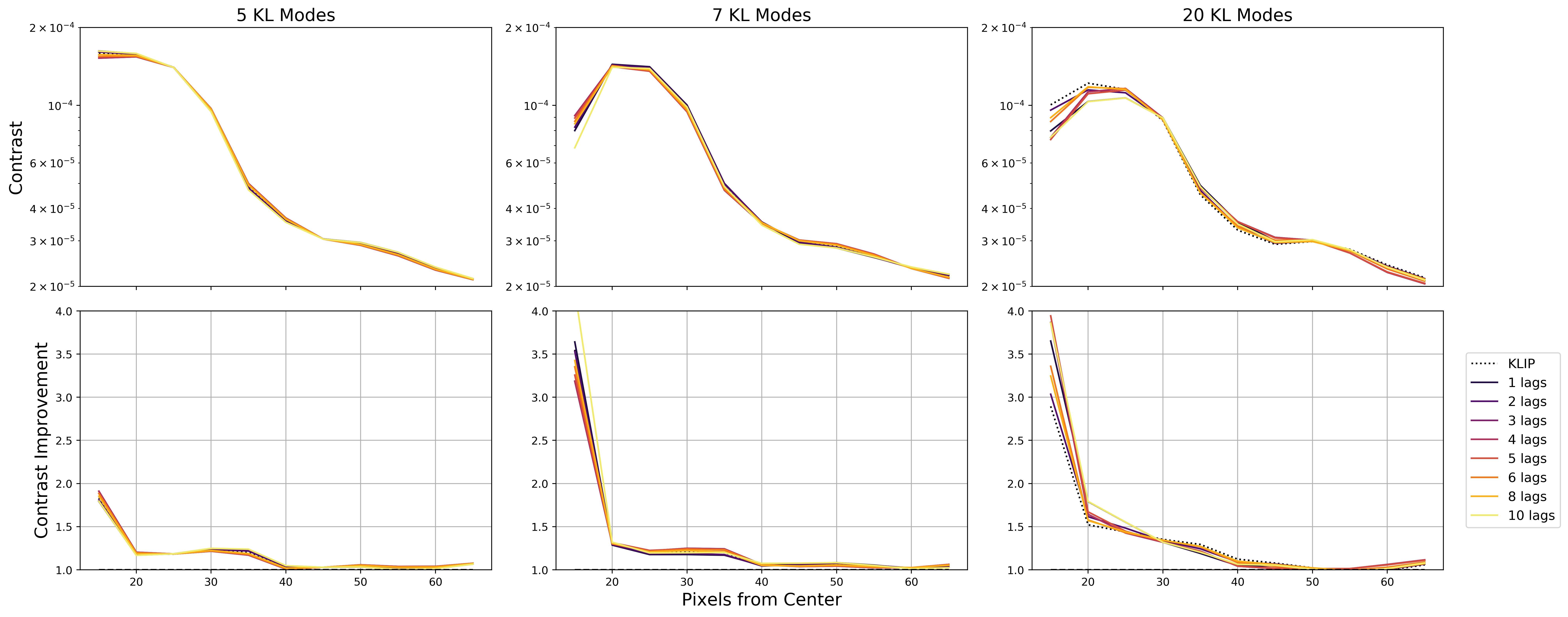}
    \caption{Contrast curves, as well as contrast improvement (a comparison to the original image's contrast curve), for three cases of KL modes: 5, 7, and 20. Each shows results for the image processed with baseline KLIP (0 lags) as well as stKLIP with a variety of lags. stKLIP is consistent with KLIP improvements, and in certain regions may show improvements depending on number of lags used.}
    \label{fig:contrast}
\end{figure*}

Contrast curves (as shown in Figure \ref{fig:contrast}) similarly show potential for up to 50\% improvement depending on the number of modes, lags, and region of the image. Within 20 pixels, we see potential for up to 400\% improvement, but with the caveat that this close to the coronagraphic mask, measurements of SNR and contrast are less reliable. A slight spread in the contrast curves for various lags, such as that seen around 30--40 pixels for 5 modes in Figure \ref{fig:contrast}, indicates that it is necessary to strategically choose the number of lags used in stKLIP depending on the image region in which we want to optimize contrast. We will discuss these results and future work further in the following section.

\section{Discussion} \label{sec:discussion}

Overall, our tests on simulated data (Section \ref{sec:results}) show that there is a demonstrated contrast gain (or equivalently, SNR improvement) of at least 10--20\% from the original image using stKLIP with fewer than 40 modes. There is also evidence that stKLIP provides a slight advantage over spatial-only KLIP for certain choices of number of lags, number of modes, and location in image. However, the real potential for this method will be unlocked when the technique is safeguarded against self-subtraction and demonstrated on real data.

In this section, we first discuss how well the signal is retained for this new algorithm, and possibilities for future improvements to better avoid self-subtraction and retain signal in Section \ref{subsec:signal}. Next, we discuss the relationship between the lag parameter and the optimized region of the target image in Section \ref{subsec:region}. Then we consider the addition of quasi-static speckles to our currently idealized, only atmospheric speckle regime in Section \ref{subsec:quasi}. Lastly, we propose other considerations for future work and implementations of this algorithm in Section \ref{subsec:future}.

\subsection{Signal Retention}\label{subsec:signal}

The signal clearly decreases beyond $\sim$5 KL modes as shown in Figure \ref{fig:signal}, indicating that we are not only subtracting from the noise but also the companion (known as self-subtraction). If we can find a way to reduce this self-subtraction and retain signal, we could potentially further improve the contrast gain. This could possibly be accomplished by masking the location of the planet or excluding regions with high spatial covariance but low temporal covariance, but further development is needed to enable this functionality. Depending on the masking implementation, this data processing method could be used for blind searches or characterization observations. In fact, it may be particularly suited to characterization observations due to the dependence on a specific image region from the nature of atmospheric speckles.

Based on previous work on LOCI (Locally Optimized Combination of Images) \citep{lafreniere2007new, marois2014gpi, thompson2021improved}, we can expect additional contrast gains once masking is implemented. Additionally, there are other techniques used for KLIP to differentiate between signal and speckles, such as angular differential imaging (ADI, \citet{2006}), spectral differential imaging (SDI, \citet{2005PASP..117..745M}), and reference differential imaging (RDI, \citet{2003EAS.....8..233M}). Similar efforts to increase the distance between the signal and the noise in the eigenimages may be useful for stKLIP. 

Additionally, when the number of lags is zero, stKLIP simply reduces to baseline KLIP \citep{soummer2012KLIP} as mentioned in Section \ref{subsec:stKLIP}, and we have included spatial-only/baseline KLIP as a comparison for stKLIP in our analyses. It is worth noting, however, that KLIP is typically used on long-exposure images, a different regime than that for which stKLIP is useful. Additionally, we are comparing stKLIP to KLIP with no self-subtraction mitigation. Most current implementations of KLIP, such as pyKLIP \citep{2015ascl.soft06001W}, do have some sort of self-subtraction mitigation or method to increase spatial diversity implemented, such as forward modeling, angular differential imaging, or spectral differential imaging \citep{pueyo2016detection,marois2006angular,vigan2010photometric}. Therefore, in practice, KLIP would currently have a significant advantage over stKLIP as implemented in this work. \rev{However, future work can adapt many of the existing methods and techniques from KLIP to improve the implementation of stKLIP and its resulting performance.}

\subsection{Optimization for Lags and Image Region}\label{subsec:region}

Despite KLIP's apparent advantages, it appears that, depending on the number of lags used and the location in the image, stKLIP can outperform KLIP by a few percent without self-subtraction implemented for either case as is done in our test. This is evident in Figure \ref{fig:contrast}, showing detail of the region with highest contrast gain (other than near the central mask). The region of highest contrast gain will vary depending on the chosen lag as well as the atmospheric conditions creating the speckles in question.  Optimization of input parameters is a notoriously tricky problem for KLIP \citep{2021AAS...23734405A}, and it appears stKLIP is subject to the same challenges. 

The variation of optimal lag and image region is due to the relationship between the wind speed and spatial frequency, since wind speed and telescope diameter combine to determine the crossing time for one cycle of the spatial frequency as $t_{{\rm cross}} = d_{{\rm telescope}} / v_{{\rm wind}}$. Spatial frequency in the pupil then corresponds to a location in the image plane. The effect of atmospheric parameters on speckle properties is further quantified in \citet{guyon2005limits} and speckle lifetimes are observed on shorter scales in \citet{2018PASP..130j4502G}. Empirical investigations of telescope telemetry and ambient weather conditions are also an ongoing area of study, especially with regards to predictive control \citep{2019AAS...23310403G, 2014SPIE.9148E..1ZR,2021arXiv211201437H}, but that information may additionally be useful in determining optimal parameters for stKLIP on-sky. \rev{Additionally, using this information on the temporal/spatial locations of strongest correlations, it may be possible to reduce the matrix size or use only the most correlated images such as in T-LOCI }\citep{marois2013tloci}.

\rev{For this work, we have been operating in the regime of milliseconds to track atmospheric speckle motions. However, in practice, the full 3D space-time correlation matrix will have power on multiple timescales, from that of atmospheric speckles to quasi-static speckles. It is outside the scope of this work to fully explore how space-time KLIP could be applied on multiple time domains, and there is additionally the caveat that computational complexity grows with longer timescales than those we have applied here.}

\subsection{Including Quasi-Static Speckles}\label{subsec:quasi}

As mentioned in Section \ref{sec:intro}, the scenario we have investigated is an idealized case --- one in which quasi-static speckles are absent and our images are dominated entirely by atmospheric speckles. We are also working on short timescales, where the atmosphere is frozen at each time step. There is a timescale over which the intensity changes, which we are observing in this scenario, but there is also a timescale for changes in the electric field's phase. These phase changes will only result in changes in intensity if superimposed onto a constant electric field, such as the case of non-coronagraphic imaging, or when quasi-static speckles are significant ($C(\vec{x})$ in Equation \ref{eq:three}). This is another regime in which to explore algorithm performance, wherein quasi-static and atmospheric speckles co-exist and interact, possibly even changing the speckle lifetimes \citep{10.1117/12.551985,2006ApJ...637..541F,2001ApJ...558L..71B,2004SPIE.5490..495S}. In this regime, there will likely be additional space-time variation as ``pinned'' speckles oscillate. Given that the presence of quasi-static speckles will make visible the additional space-time variations in phase, it is possible that stKLIP will operate even more effectively with this additional information to exploit. 
However, additional quasi-static speckles will lead to additional photon noise, which may counteract any theoretical contrast gains from including phase information. (Note: recent work from \citet{2019AAS...23314038M} shows that using KLIP on shorter exposures may even help remove quasi-static speckles more effectively, further bolstering the case for the stKLIP's effectiveness in this regime.) Additionally, the presence of atmospheric residuals could even provide information about the phase of quasi-static speckles, allowing them to be effectively nulled with a deformable mirror \citep{10.1117/12.2054356,Frazin:21}. Future simulations may explore this regime and determine if additional contrast gain is possible.

\subsection{Considerations for Future Work}\label{subsec:future}

In this idealized test case, we also chose not to simulate adaptive optics corrections, instead leaving an investigation of how AO parameters affect space-time correlations and the resulting stKLIP processing for a future investigation. Since AO suppresses low frequencies and heaves high frequencies unchanged, although our total error is on par with an AO residual scenario, the overall shape of the power spectrum would be different. This would likely lead to weaker temporal correlations with AO. Previous work also shows that AO corrections do affect the lifetime of speckles \citep{2021PASP..133j4504M, males2017groundbased}, so this will be an important factor to consider in future work.

Currently, we have yet to demonstrate the full potential of this algorithm, in part due to the high computational costs. To run stKLIP on a 100$\times$100 pixel window of a simulated 30-second data set (with the parameters specified in Section \ref{sec:results}) over a range of KLIP parameters, we required 128GB of RAM and approximately 400 hours of computation time. The high memory requirement is due to the eigendecomposition, since the space-time covariance matrix can become extremely large when including a large number of lags and must be loaded in fully to the eigendecomposition. As mentioned in Section \ref{subsec:iterative}, there are possible solutions to this challenge to reduce computational costs in less memory intensive implementations, or even analytical gains in efficiency that exploit symmetries inherent in the covariance matrix (shown in Figure \ref{fig:stKLIP-cartoon}) \rev{or focus on only the strongest correlations depending on the temporal and spatial scales of interest}, but those are beyond the scope of this paper. 

It may also be possible to reduce the number of eigenvalues/modes computed, which will reduce computation time and possibly memory consumption as well, given that we now know that values beyond $\sim$50 KL modes aren't of much use in our tested scenario, but the exact threshold will be dependent on the region of interest and number of lags used, among other factors. In future iterations, this code could also likely be improved by implementing this algorithm more optimally rather than in a high-level language, as the current implementation is in Python, and by using parallel processing.

\section{Conclusion} \label{sec:conclusions}

Evolving atmospheric layers lead to time-varying speckles in the focal plane of an imaging system; for the high-contrast imaging regime, we have shown that spatio-temporal covariances in these speckles exist, and can be exploited for use in data processing to improve contrast. Our data processing tool has been implemented in Python, tested on a simple analytic test case to prove viability, and also tested on realistic simulations to understand the effectiveness of this technique. We have shown there is potential for a contrast gain (or equivalently, SNR improvement) of at least 10--20\% from the original image, with significant potential for an even larger gain if self-subtraction is adequately addressed. Additionally, we have shown evidence that the space-time nature of our algorithm, in its current form, may provide a slight advantage over spatial-only KLIP in certain cases, with significant potential for stronger improvement under different conditions and with improvements to the algorithm implementation. Although the SNR gains for this new method aren't fully developed, this initial work on space-time KLIP opens the door for the use of space-time covariances in high-contrast imaging, especially in the short timescale regime of atmospheric speckle lifetimes.

Future work can use our data processing tool to further explore the dependence of the space-time covariances and the resulting contrast improvements on various parameters, such as the type of coronagraph, AO performance, strength of quasi-static speckles, and atmospheric conditions. It would be particularly interesting to determine how AO affects these covariances, since AO is important in a realistic scenario for exoplanet imaging and affects the resulting speckle lifetimes and structures. 

Future implementations of this algorithm will also need to consider how to minimize self-subtraction of the companion object, and overcome the memory and computational demands in the eigendecomposition. Further optimization of the tunable parameters is also necessary to optimize algorithm performance and implement this as a refined tool for exoplanet imaging. It would also be interesting to apply this tool to on-sky data, such as that from MEC on SCExAO at Subaru \citep{walter2020mkid, walter2018mec, 2015PASP..127..890J, 2010SPIE.7736E..3NM}, to determine potential on-sky contrast gains from this technique. Although this current work focuses on the use of speckle space-time covariances in post-processing, these covariances could even be used in real-time predictive control \citep{2018SPIE10703E..0ZG}. Overall, the results in this work show that harnessing space-time covariances through ``space-time KLIP'' may be a promising technique to add to our toolkit for suppressing speckle noise in exoplanet imaging while retaining signal throughput.

\hspace{5cm}

\begin{acknowledgements}

This work used computational and storage services associated with the Hoffman2 Shared Cluster provided by UCLA Institute for Digital Research and Education’s Research Technology Group. This work was supported in part by National Science Foundation award number 1710514 and by Heising-Simons Foundation award number 2020-1821. This material is based upon work supported by the National Science Foundation Graduate Research Fellowship under Grants No. 2016-21 DGE-1650604 and 2021-25 DGE-2034835. Rupert Dodkins is supported by the National Science Foundation award number 1710385. Kristina K. Davis is supported by an National Science Foundation Astronomy and Astrophysics Postdoctoral Fellowship under award AST-1801983. Any opinions, findings, and conclusions or recommendations expressed in this material are those of the authors(s) and do not necessarily reflect the views of the National Science Foundation. Thanks to Marcos M. Flores and Joseph Marcinik for helpful discussions on notation and LaTeX.

\end{acknowledgements}

\hspace{5cm}

\software{NumPy \citep{numpy},
IPython \citep{ipython}, Jupyter Notebooks \citep{jupyter},
Matplotlib \citep{matplotlib}, 
Astropy \citep{astropy:2013, astropy:2018}, SciPy \citep{scipy}, h5py \citep{hdf5}, MEDIS \citep{dodkins2018development,dodkins2020medis}, Dask \citep{dask,matthew_rocklin-proc-scipy-2015}}

\hspace{5cm}

\bibliography{speckle-bib}

\appendix
\section{Notation Glossary -- Sections 2.1 \& 2.3}\label{var-table1}
\begin{tabular}{|p{2cm}|p{8cm}|}
\hline
\textbf{Symbol} & \textbf{Definition} \\ 
\hline
 $I_\psi(k)$ & Stellar PSF, as in \citet{soummer2012KLIP}\\
 $k$ & Pixel index \\
 $T(k),\vec{t}$ & Target image as in \citet{soummer2012KLIP}, and as in this work unrolled to 1-d and represented as a vector \\
 $A(k), \vec{a}$ & Faint astronomical signal (as above) \\
 $\epsilon$ & True/false binary parameter \\
 $\hat{I_\psi}(k), \vec{\hat{\psi}}$ & Approximated PSF as in \citet{soummer2012KLIP} and as a vector in this work, respectively\\
 $R$ & Matrix of reference images before mean subtraction \\
 $\vec{r}$ & Individual reference image \\
 $X$ & Mean image from reference set \\
 $M$ & Mean subtracted reference images \\
 $n_{i}$ & Number of images in reference set \\
 $n_{p}$ & Pixel count $n_x \times n_y$ \\
 $n_x$ & Dimension 1 size \\
 $n_y$ & Dimension 2 size \\
 $n_m$ & Number of modes / eigenvectors chosen \\
 $i$ & Used as an arbitrary index \\
 $j$ & Used as an arbitrary index \\
 $C$ & Covariance matrix \\
 $\vec{\lambda}$, $\lambda$ & Vector of eigenvalues, eigenvalue \\
 $V$ & Matrix of eigenvectors/eigenimages \\
 $\vec{v}$ & Eigenvector \\
 $\vec{q}$ & Vector of coefficients \\
 $S$, $\vec{s}$ & Mean subtracted image sequences (in matrix and vector form) \\
 $\vec{\hat{s}}$ & Reconstructed image sequence \\
 $n_s$ & Number of images in sequence \\
 $L$ & Largest number of timesteps/lags in use as measured from the central image \\
 $n_l$ & Total number of timesteps/lags used, equal to $n_s$\\
 $\vec{r}_{k,avg}$ & Averaged residual from stKLIP \\
 \hline
\end{tabular}

\section{Notation Glossary -- Section 2.2}\label{var-table2}
\begin{tabular}{|c|c|}
\hline
\textbf{Symbol} & \textbf{Definition} \\ 
\hline
 $I$ & Intensity \\  
 $\vec{x}$ & Location in image plane \\
 $\vec{u}$ & Location in pupil plane \\
 $t$ & Time \\
 $\tau$ & Time step \\
 $\Psi_{pup}$ & Pupil amplitude \\
 $\Psi_{foc}$ & Focal amplitude \\
 $\psi(\vec{u},t)$ & Pupil phase \\
 $P(\vec{u})$ & Pupil function \\
 $C(\vec{x})$ & Spatially coherent wavefront \\
 $S_\phi(\vec{x},t)$ & Phase aberrations \\
 $\xi$ & Displacement in pupil \\
 $B_\phi$ & phase covariance function \\
 $v_{wind}$ & Wind velocity \\
 \hline
\end{tabular}

\end{document}